\RequirePackage{fix-cm}
\documentclass{svjour3}                     
\smartqed  
\usepackage{graphicx}
\usepackage{amsmath,amssymb,amsfonts}
\usepackage{multirow}
\usepackage{amsmath,amssymb,amsfonts}
\usepackage{algorithm}
\usepackage{graphicx}
\usepackage{textcomp}
\usepackage{xcolor}
\usepackage[utf8]{inputenc}
\usepackage{listings}
\usepackage{float}
\usepackage{subfigure}
\usepackage{url}
\usepackage{multirow}
\usepackage{commath}
\usepackage{algpseudocode}
\usepackage{etoolbox}
\usepackage{varwidth}
\usepackage{tikz}
\usetikzlibrary{shapes,arrows}
\usepackage{colortbl}
\usepackage{appendix}
\usepackage{multicol}
\usepackage{cite}

\definecolor{codegreen}{rgb}{0,0.6,0}
\definecolor{codeblack}{rgb}{0.30,0.30,0.30}
\definecolor{codepurple}{rgb}{0.58,0,0.82}
\definecolor{backcolour}{rgb}{0.95,0.95,0.92}
 
\lstdefinestyle{mystyle}{
    backgroundcolor=\color{backcolour},   
    commentstyle=\color{codegreen},
    keywordstyle=\color{magenta},
    numberstyle=\scriptsize\color{codeblack},
    stringstyle=\color{codepurple},
    breaklines=true,
    breakatwhitespace=true,
    numbers=left,
    xleftmargin=0.2in,
    xrightmargin=0.05in,
    basicstyle=\ttfamily\footnotesize,
    numbersep=5pt,
    frame=single,                  
    tabsize=1
}

\journalname{ArXiv Preprint}
\begin{document}

\title{PPT-Multicore: Performance Prediction of OpenMP applications using Reuse Profiles and Analytical Modeling}

\titlerunning{PPT-Multicore}        

\author{
Atanu Barai\and Yehia Arafa\and Abdel-Hameed Badawy\and \\
Gopinath Chennupati\and Nandakishore Santhi\and Stephan Eidenbenz
}

\authorrunning{Barai, Arafa, Badawy, \textit{et. al.}} 

\institute{Atanu Barai 
            \at New Mexico State University.
            \email{atanu@nmsu.edu}
            \and
            Yehia Arafa  
            \at New Mexico State University.
            \email{yarafa@nmsu.edu}
            \and
            Abdel-Hameed Badawy 
            \at New Mexico State University.
            \email{badawy@nmsu.edu}
            \and
            Gopinath Chennupati 
            \at Los Alamos National Laboratory.
            \email{cgnath.dr@gmail.com}
            \and
            Nandakishore Santhi 
            \at Los Alamos National Laboratory.
            \email{nsanthi@lanl.gov}
            \and
            Stephan Eidenbenz
            \at Los Alamos National Laboratory.
            \email{eidenben@lanl.gov}
}

\date{}

\def\makeheadbox{\relax}
\maketitle

\begin{abstract}
We present \emph{PPT-Multicore}, an analytical model embedded in the Performance Prediction Toolkit (PPT) to predict parallel application performance running on a multicore processor. \emph{PPT-Multicore} builds upon our previous work towards a multicore cache model. We extract LLVM basic block labeled memory trace using an architecture-independent LLVM-based instrumentation tool only once in an application's lifetime. The model uses the memory trace and other parameters from an instrumented sequentially executed binary. We use a probabilistic and computationally efficient reuse profile to predict the cache hit rates and runtimes of OpenMP programs' parallel sections. We model Intel's Broadwell, Haswell, and AMD's Zen2 architectures and validate our framework using different applications from PolyBench and PARSEC benchmark suites. The results show that \emph{PPT-Multicore} can predict cache hit rates with an overall average error rate of 1.23\% while predicting the runtime with an error rate of 9.08\%.

\end{abstract}

\section{Introduction}
\label{sec:intro}
\vspace{-5pt}
Nowadays, with the emergence of Exascale computing, multicore processors with hundreds of cores, complicated memory hierarchy, instruction pipelining, branch prediction, and aggressive speculative execution have been a standard rather than an exception, from mobile devices to supercomputers. Such complicated designs come with several challenges~\cite{john:exascale}, such as the efficient use of available computing cycles and the percentage of available memory utilization. Software designers have to fully leverage modern processors' extensive computing power, especially the parallelization primitives. One of the critical factors determining a parallel application's performance on a multicore is the data availability to the computing units. One way to measure an application's data availability is through its cache utilization ability, which directly impacts runtime performance.

Modeling and simulation (ModSim) tools have been used to a great extent to help in understating the limiting factors and bottlenecks on applications' performance. Co-design, which we define as modeling both hardware and software, helps tune an application's performance. Most of the efforts in co-design have focused on getting simulation data from cycle-accurate dynamic instrumentation tools~\cite{Davis-max-cmp-thpt,Huh-des-sp,Moguls}. However, these simulations require a large number of runs and experimentation with many hardware configurations. Such configurations include variations in cache hierarchies, core counts, and problem sizes, all of which contribute to increasing design space complexity. Using cycle-accurate dynamic simulators to model and predict performance does not scale well and usually needs days, if not weeks and months, to produce output for large-scale applications. This slowdown is the huge bottleneck of the simulation tool. A co-design framework that is accurate in terms of prediction and scalable in terms of simulation time and core count is crucial in analyzing a multicore system's performance.

In this paper, we introduce the \emph{PPT-Multicore}. It combines and extends concepts from the Scalable Analytical Shared Memory Model (SASMM)~\cite{ppt-sasmm} and the Performance Prediction Toolkit (PPT)~\cite{ppt}. \emph{PPT-Multicore} is a performance model based on code analysis and reuse distance (the number of unique references between two references to the same address~\cite{Mattson:RD:IBM}) estimation methods. PPT is a parameterized co-design framework developed to predict application runtime at Los Alamos National Laboratory. PPT relies on a high event rate Parallel Discrete Event Simulation (PDES) engine named Simian~\cite{simian},  written in Python, Lua, and JavaScript. Simian allows mixing both entity and process-oriented simulation models and conservative, optimistic, or hybrid operation modes.

\emph{PPT-Multicore} estimates the shared and private cache hit rates and overall runtime of parallel sections of an OpenMP~\cite{openmp} application running on a multicore architecture. The main building blocks of \emph{PPT-Multicore} are the following: (i) We translate the parallel sections of the input OpenMP application to threaded version code using the {\em Rose} compiler~\cite{Rose_Compiler:Liao}. (ii) We collect LLVM basic block~\cite{Lattner:LLVM} labeled memory trace of the parallel sections from a sequential execution of translated code. (iii)  Using this memory trace, we explore different scheduling and interleaving strategies to mimic the behavior of multi-threaded programs on shared-memory multicores. We carry out these strategies at the basic block level. (iv) We mimic traces for the private and shared caches from the sequential trace and apply a probabilistic analytical method to measure the reuse distance profiles. Using these profiles, we estimate cache hit rates of the parallel sections of OpenMP applications. (v) We then pass the hit rates and other application-specific information such as the number of different arithmetic operations and the total memory operations to the \emph{PPT-Multicore} Simian PDES model for runtime predictions.

To evaluate our tool, we compare the predicted cache hit rates with hit rates collected from real hardware using PAPI~\cite{papi-c}. We also evaluate the predicted runtime by comparing it with timings from runs on real hardware. The results show that the model accurately predicts cache hit rates and runtimes across a large set of benchmark applications. Our model can predict cache hit rates with an overall average accuracy of 98.77\%, while for runtime prediction, the accuracy is 90.92\%.

The contributions of the paper can be summarized as follows:
\vspace{-1ex}
\begin{itemize}
    \item Vastly improved memory model for private and shared caches' hit rate predictions on multicores    
\item Accurate runtime prediction of parallel sections of OpenMP applications
    \item Predicting the performance of OpenMP applications from the single thread execution trace of the application
    \item Showing the predictions for various core counts without having to rerun the application.
\end{itemize}

The rest of the paper is organised as follows: Section~\ref{sec:background} briefly discusses the OpenMP execution model and introduces reuse distance theory for both single and multicore processors; Section~\ref{sec:method} describes our modeling approach; In Section~\ref{sec:results}, we evaluate our model and describe the results; While Section~\ref{sec:related_work} discusses relevant related works, and finally Section~\ref{sec:conclusion} concludes the paper.
\section{Background}
\label{sec:background}
\subsection{Fork Join Model: Execution of Parallel Application}
In OpenMP shared-memory programming, a program exploits parallelism using the fork-join model. Program execution begins sequentially with only the primary thread. When it encounters a parallel region pragma, the primary thread forks a team of sub-threads. By default, the sub-threads execute the code in the parallel sections independently and can access all the variables declared in the primary thread before the fork. These variables are referred to as shared variables. A programmer can also specify private variables for each thread. When the threads complete a parallel section, they synchronize and join; the program execution continues with the primary thread.
\subsection{Performance Modeling of Parallel Applications}
Parallel applications exploit multicores with hundreds of cores. Thus, it is important that the programs can utilize all the hardware resources available to increase performance. We can obtain valuable insight into a parallel application's performance on real hardware by developing their performance model. Based on HW/SW co-design, performance modeling helps improve the understanding of the applications' behavior and modify for best performance. It also helps us study the impact of advanced architectural designs and features on performance in multicores and manycores. Therefore, we can tune an application for the best performance. We can also suggest optimal hardware for a group of applications by simulating different hardware architectures' performance. Thus, performance modeling plays a vital role in parallel program design.

On the other hand, performance modeling and evaluation are at the heart of modern parallel computer architecture research and development. Because of the growing complexity of modern processors, architects cannot design systems based only on intuition~\cite{survey_evoulution_methods}. Rigorous performance evaluation methodologies are a crucial part of modern architecture research and development.

\newcolumntype{P}[1]{>{\centering\arraybackslash}m{#1}}
\begin{table}[t]
    \centering
    \caption{Reuse Distance Example}
        \begin{tabular}{|c|cccccccc|}
            \hline
            Access Time & 1 & 2 & 3 & 4 & 5 & 6 & 7 & 8 \\
            \hline
            Memory Address & w & x & w & y & x & z & z & w\\
            \hline
            {\text { Reuse Distance }} & {$\infty$} & {$\infty$} & {1} & {$\infty$} & {2} & {$\infty$} & {0} & {3}\\
            \hline
        \end{tabular}
    \label{fig:RD_theory}
\end{table}

\begin{table}[t]
    \centering
    \caption{Reuse Profile Example}
    \begin{tabular}{|c|ccccc|}
        \hline
        Reuse Distance & 0 & 1 & 2 & 3 & $\infty$\\
        \hline
        Frequency/Count & 1 & 1 & 1 & 1 & 4 \\
        \hline
        P(RD) & {1\(/ 8\)} & {1\(/ 8\)} & {1\(/ 8\)} & {1\(/ 8\)} & {4\(/ 8\)} \\
        \hline
    \end{tabular}
    \label{fig:RD_profile}
\end{table}
\vspace{-10pt}
\subsection{Reuse Distance}
Reuse distance (D) of a memory address, is denoted as the number of unique memory references between two consecutive references to the same address. It is also known as the LRU stack distance~\cite{Mattson:RD:IBM}. If a memory address is accessed for the first time, the reuse distance D for that access is $\infty$. Table~\ref{fig:RD_theory} shows an example of reuse distance calculation for a sample memory trace. For the access of address \emph{w} at time \emph{8} the RD~\footnote{D and RD are used interchangeably} value is 3 as there are three unique memory references from its previous access at time 3.

Reuse profile is the histogram of reuse distances for all memory references of a memory trace. Table~\ref{fig:RD_profile} shows reuse profile for the example memory trace shown in Table~\ref{fig:RD_theory}. In the histogram, D sits on the X-axis while Frequency or probability of D (\emph{P(D)}) sits on the Y-axis. Reuse distance analysis can be used to measure locality~\cite{locality:Ding:2003:PWL,locality:Zhong:2009:PLA} which in turn can be used to predict the cache performance of that application~\cite{performance:Beyls:RD:2001,performance:CaBetacaval:2003:ECM,performance:Sen:2013:ROM} and make cache management policy decisions~\cite{C.Management:Duong:2012:ICM}. Furthermore, reuse distance analysis has also been used for parallel application performance prediction for compile-time optimizations or design space explorations~\cite{badamo-ml-rd-power,badawy-cal,badawy-ipccc,wu-msp,minshu-ml-rd}. We estimate cache misses based on the 3C model; compulsory, capacity, and conflict misses. For a fully associative cache with capacity C, a memory reference's reuse distance will always trigger a cache miss if D $\geq$ C. In the example shown in Table~\ref{fig:RD_theory}, $50\%$ of memory references with RD value of $\infty$ will cause a compulsory cache miss. If we consider that cache size is four, then none of the memory references will cause a capacity cache miss.

Reuse distance analysis is robust and architecture-independent for sequential applications. Once the memory trace of an application is collected and a reuse profile generated, we can compute different cache configurations' performance. It saves a significant amount of time in cache performance analysis and design as we do not have to collect memory traces for different cache configurations. Previous attempts~\cite{Berg:SS,Ding:2001:RDA,Steen:UGhent} demonstrated the use of memory traces for reuse profile calculations. These approaches use binary instrumentation tools to collect memory traces. The memory traces used in most of these attempts are significant in size and time-consuming to process, thereby unscalable. However, recent attempts from Chennupati \textit{et al.}~\cite{chennupati:pmbs,chennupati:pads,ppt-amm} demonstrated analytical models that scale with a small input run of a program. These attempts help predict the performance of an application on single-threaded programs. Similarly, we model the private and shared cache performance of multicore programs using reuse distance for multicores. We discuss Reuse Distance Analysis on Multicores in the next Section~\ref{Subsection:RD-Multicore}.
\vspace{-15pt}
\subsection{Reuse Distance Analysis on Multicore Processors}
\label{Subsection:RD-Multicore}
Most multicores have multiple cache hierarchies with private and shared caches. A core accesses its private cache while all the cores access the shared cache. Although the locality of references of a parallel program running on a multicore is somewhat architecture-specific, it largely depends on the application's memory access characteristics. Two separate reuse profiles, a \textit{Private-stack} (PRD) and \textit{Concurrent} reuse profiles (CRD) are used to model private and shared caches~\cite{Jiang:RD-Applicable-on-chip} respectively. On a shared cache, memory references from different cores interleave with one another. Thus, we can interleave memory references from different cores on a single LRU stack to measure concurrent reuse profiles. This interleaving causes different types of interaction: \textit{dilation, overlap, and interception}~\cite{Wu-multicore-journal}.

Table~\ref{fig:CRD_theory} shows the memory references from two cores. Reference \textbf{u} at time 4 has a CRD of two, while its PRD is 1. Here, the CRD is larger than the PRD, which shows \textit{dilation}. On the other hand, data sharing reduces dilation. Reference \textbf{u} at time 9 has a CRD of three, although there are four memory references between the two consecutive memory references to \textbf{u} at times 4 and 9. This shows \textit{overlapping} as \textbf{x} is accessed by both cores inside the reuse interval of \textbf{u}. For \textbf{v} at time 10, the reused data itself is shared. Thus, its CRD is two, which is less than its PRD.

\begin{table}[t]
    \centering
    \caption{Concurrent Reuse Distance Example}
    \label{fig:CRD_theory}
    \begin{tabular}{|P{2.5cm}|cccccccccc|}
        \hline
        Time       & 1 & 2 & 3 & 4 &  5 &  6 & 7 & 8 & 9 & 10\\
        \hline
        Core $C_1$ & u &   & v & u &  y &  &   & x & u & v\\
        \hline
        Core $C_2$ &   & w &   &   &   & x & v &   &   &  \\
        \hline
        Shared Memory Access & u  & w & v  & u  & y  & x & v & x  & u  & v \\
        \hline
    \end{tabular}
\end{table}

Several recent works have focused on CRD profiles for predicting the performance of shared cache~\cite{Multicore:Formalizing_Data_Locality:Ceballos,Multicore:Performance_metrics:Ding2014,Multicore:Modeling_CMP_Cache_Capacity:Shi,Multicore:Modeling_Superscalar_Memory-Level_Parallelism,Multicore:Miss_Rate_Prediction:Zhong}. Recently, researchers attempted to use analytical model and sampling to speed up the performance prediction~\cite{multicore:stat_multiprocessor_cache:Berg,Jiang:RD-Applicable-on-chip,Multicore_Reuse_Analytical:Jasmine,Schuff:2010:AMR:1854273.1854286,Multicore-Aware-Derek}. All these models require trace collection from parallel executions of the application for different thread counts. On the other hand, our model collects a trace once from a sequential run of the application. From the sequential trace, we predict the shared cache performance for different thread counts. This makes our model scalable in terms of core count.
\section{Methodologies}
\label{sec:method}

\begin{figure*}[t]
      \centering
      \includegraphics[width=0.98\linewidth]{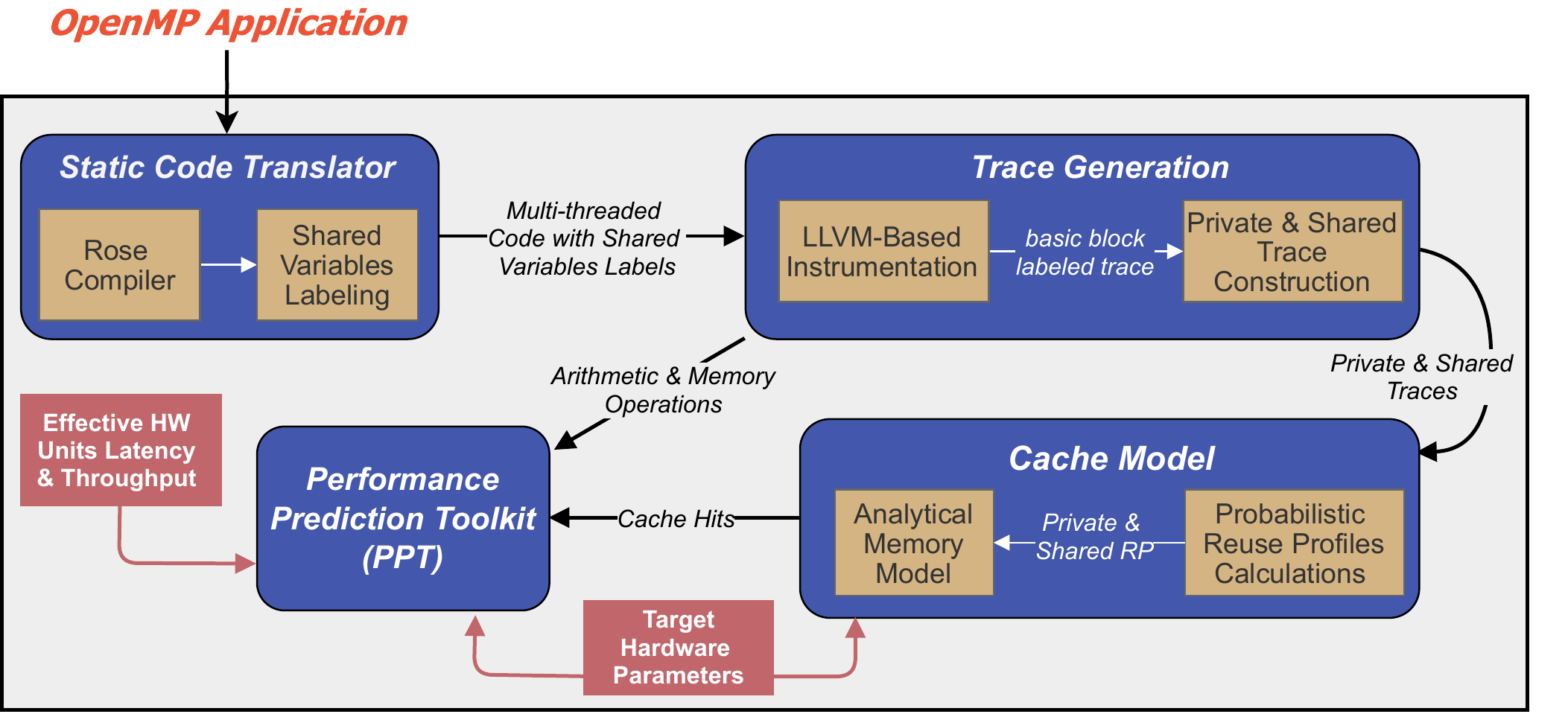}
      \caption{Overview of PPT-Multicore}
      \vspace{-10pt}
      \label{fig:flowchart}
\end{figure*}

\emph{PPT-Multicore} is a parameterized tool for performance prediction of a parallel OpenMP code. We leverage reuse distance analysis to determine the \textit{Private-stack} (PRD) and \textit{Concurrent} reuse profiles of the parallel sections of a program. These reuse profiles are then used to determine the hit rates of different cache levels and, ultimately, the parallel sections' runtime. Figure~\ref{fig:flowchart} shows the different steps of our model that include \emph{a)} translating the OpenMP code to a threaded code and adding labels for shared variables in the threaded program, \emph{b)} generating a basic block labeled memory trace from a sequential run and mimicking shared, and private memory traces from it, \emph{c)} estimating private and concurrent reuse profiles and hit rates, and \emph{d)} calculating the effective latency, throughput, and finally predicting runtime. In this section, We describe each of these steps in detail.

\begin{figure}[b]
    \centering
    \lstset{style=mystyle}
    \begin{lstlisting}[language=C,numbers=left]
int  main()
{
  int i, n, sum;
  n = 500;
  sum = 0;
#pragma omp parallel for reduction(+: sum)
  for(i = 0; i < n; i++)
    sum = sum + i;
}
\end{lstlisting}
\caption{A simple OpenMP summation program}
\label{fig:simple_openmp}
\end{figure}

\begin{figure}[htbp]
\centering
\lstset{style=mystyle}
\begin{lstlisting}[language=C,,numbers=left]
//Structure for shared variables
struct shared_struct {
  void *n_p;
  void *sum_p;
};

int main(int argc,char **argv) {
  int status = 0;
  XOMP_init(argc,argv);
  int i, n, sum;
  n = 500;
  sum = 0;
  struct shared_struct shared_var;
  shared_var.sum_p = ((void *)(&sum));
  shared_var.n_p = ((void *)(&n));
  XOMP_parallel_start(OUT__1__7285__,&shared_var);
  XOMP_parallel_end();
  XOMP_terminate(status);
}

static void OUT__1__7285__(void *s_data) {
shared_var_trace0: {} //Label for shared variables
  int *n = (int *)(((struct shared_struct *)s_data)->n_p);
  int *sum = (int *)(((struct shared_struct *)s_data)->sum_p);
other_trace1: {} //Label for rest of the program
  int _p_sum = 0;
  long index, lower, upper;
  XOMP_loop_default(0, *n - 1,1, &lower, &upper);
  for (index = lower; index <= upper; index += 1)
    _p_sum = _p_sum + index;
  XOMP_atomic_start();
   *sum =  *sum + _p_sum;
  XOMP_atomic_end();
  XOMP_barrier();
}
\end{lstlisting}
\caption{Example of how the code in figure~\ref{fig:simple_openmp} is translated to intermediate threaded code using Rose compiler. Note how the parallel section in figure~\ref{fig:simple_openmp} has been converted to function \textit{$OUT\_\_7285\_\_$}. A structure of pointers to the shared variables is passed to the function. Labels are added for identifying shared variables.}
\label{fig:simple_rose_code}
\end{figure}
\vspace{-10pt}
\subsection{Static Code Translation}
\label{sec:prog_translation}
In the first step of \emph{PPT-Multicore}, we convert the OpenMP application to an intermediate threaded code using the OpenMP translator in ROSE~\cite{Rose_Compiler:Liao} compiler framework. . In the translation process, the original code's parallel sections are transformed into intermediate functions to be executed parallelly. The shared variables accessed by the parallel sections are also grouped in the intermediate code. Tracking the memory references for the shared variables is difficult in the high-level OpenMP code. Therefore, the translation is an important step to correctly track the shared and private variables in the parallel regions. Thus, the translated code helps to calculate PRD and CRD profiles efficiently. Furthermore, the names of these intermediate functions corresponding to the parallel sections start with \emph{`OUT\_\_'}, making it easier to generate memory traces only for the parallel sections of the program. We only need to instrument those intermediate functions that are hard to complete in the high-level OpenMP code.


Figure~\ref{fig:simple_openmp} and ~\ref{fig:simple_rose_code} show a simple OpenMP program and its translated intermediate threaded code respectively. The translated code contains XOMP wrapper functions generated from the Rose compiler. These wrapper functions call the GNU OpenMP (GOMP) library functions internally when compiled with GCC. The function named \emph{OUT\_\_1\_\_7285} corresponds to the OpenMP code's parallel section in the translated code. The private variables of the OpenMP parallel sections are translated as local variables of \emph{OUT\_\_1\_\_7285}. As each thread under execution runs its copy of these functions, memory allocation for local variables is also done individually.

The translated functions in the threaded version of the code receive pointers' structure as parameters (\emph{s\_data} in the example code) for the shared variables. These pointers point to the shared variables accessed by the functions. All members of these structures are assigned to locally declared pointers (line 21-22 in figure~\ref{fig:simple_rose_code}). Using a script, We put the assignment statements of shared variables under a label(see line 20 where \emph{shared\_var\_trace0} label is added). In the memory trace, all the references under the \emph{shared\_var\_trace0} label are grouped in the corresponding basic block label so that later we can identify the memory references for shared variables. We further discuss our trace construction approach in section~\ref{sec:mem_trace}.
\vspace{-10pt}
\subsection{Memory Trace Generation for Different Cache Hierarchies}
\label{sec:mem_trace}
In the second step of our model, we generate LLVM basic block labeled memory trace only for the translated threaded program's parallel functions. Each of the LLVM basic blocks has a single entry and a single exit point of execution. We can obtain the basic blocks' labels in an LLVM intermediate representation file. We use a modified version~\cite{chennupati:pmbs} of LLVM based instrumentation tool, Byfl~\cite{Byfl} which can instrument the preferred functions (in this case, functions starting with \emph{OUT\_\_}) to generate the basic block labeled memory trace through sequential execution. We add the basic block labels in the memory trace in a way that all the memory addresses that are accessed as a result of executing the corresponding straight-line code of \textit{(BB\textsubscript{i})} are grouped together. 

\begin{figure}
    \centering
    \includegraphics[width=\linewidth]{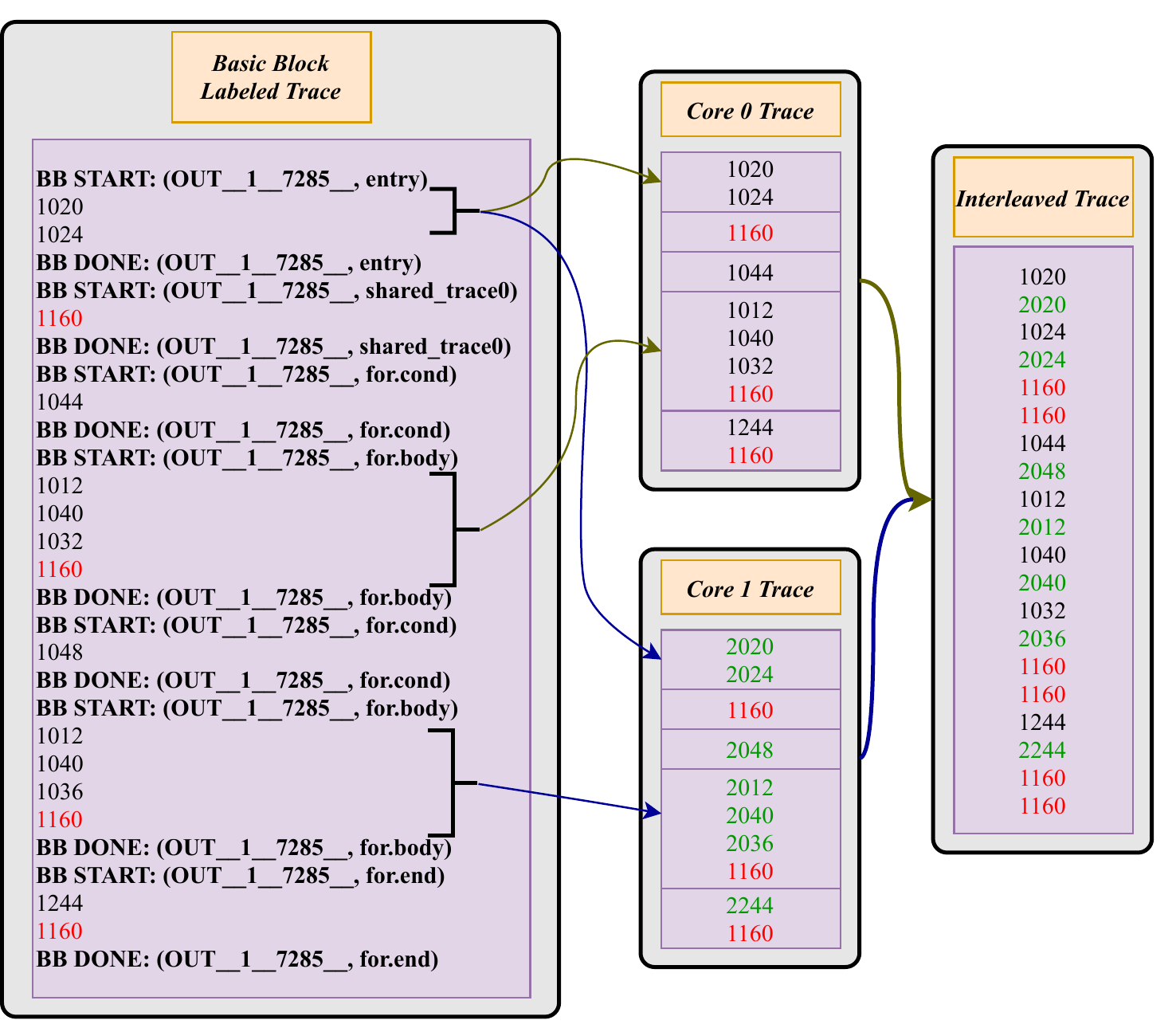}
    \caption{Example of private and shared memory trace construction from basic block labeled memory trace. For the sake of the example, we are considering only two cores and round-robin interleaving. Here, basic blocks with a single instance (\emph{e.g.}, $OUT\_\_7285\_\_,entry$) in the original trace are copied to each core's trace. On the other hand, basic blocks with instances greater than or equals the number of cores (\emph{e.g.}, $OUT\_\_7285\_\_,for.body$) are being distributed to multiple cores. While constructing core1's trace, we are adding its offset to the memory addresses (green color) except for the addresses for the shared variables (red color))}
    \label{fig:trace_construction}
    \vspace{-15pt}
\end{figure}

Figure~\ref{fig:trace_construction} shows an example of how the private and shared memory traces are constructed from the labeled basic block trace. We inspect the shared basic blocks from the identification of the basic block's label name. For instance, We gather all the memory references under the shared label (\emph{shared\_var\_trace} was added from the previous step). We then mimic the parallel section of the program's memory access behavior while running on multiple cores using the basic block labeled memory trace and thus generate the private memory traces of each thread under execution.


\begin{algorithm}[tbp]
    \begin{algorithmic}[1]
    \Procedure{$gen\_private\_traces$}{$bb\_list$, $num\_cores$, $trace$, $shared\_var\_refs$, $bb\_counts$}
        \State $private\_traces \gets [[]*num\_cores]$
        \State $bb\_count\_per\_core \gets [[]*num\_cores]$
        \State $bb\_done\_count \gets [[]*num\_cores]$
        \For{$bb\_id$ \textbf{in} $bb\_list$}
            \If{$bb\_count[bb\_id]$ $<$ $num\_cores$}
                \State $bb\_count\_per\_core[bb\_id] \gets 1$ \textcolor{codegreen}{\Comment{Each core gets a copy of BB}}
            \Else
                \State $bb\_count\_per\_core[bb\_id] \gets bb\_counts[bb\_id]/num\_cores$ \textcolor{codegreen}{\Comment{Split BB instances to different cores}}
            \EndIf
        \EndFor
        \For{$entry$ $\textbf{in}$ $trace$}
            \If{$entry$ $\textbf{is}$ $mem\_address$}
                \State $address = strtoull(entry)$
                \If{$bb\_count\_per\_core[current\_bb\_id] = 1$}
                    \For{$core = 0; core < num\_cores; core{+}{+}$} 
                        \If{$address \notin shared\_var\_refs$}
                            \State $private\_traces[core].append(address + offset \times core)$ \textcolor{codegreen}{\Comment{Add offset if the address is not for shared variable}}
                        \Else
                            \State $private\_traces[core].append(address)$
                        \EndIf
                    \EndFor
                \Else
                    \State $core \gets bb\_done\_count[bb\_id] / bb\_count\_per\_core[bb\_id]$
                    \If{$address \notin shared\_var\_refs$}
                        \State $private\_traces[core].append(address + offset \times core)$  \textcolor{codegreen}{\Comment{Add offset if the address is not for shared variable}}
                    \Else
                        \State $private\_traces[core].append(address)$
                    \EndIf
                \EndIf
            \ElsIf{$BB\_START$ $\textbf{in}$ $entry$}
                \State $current\_bb\_id \gets entry.split(: )[2]$
            \ElsIf{$BB\_END$ $\textbf{in}$ $entry$}
                \State $bb\_done\_count[bb\_id] = bb\_done\_count[bb\_id] + 1$
            \EndIf
       \EndFor
    \EndProcedure
    \end{algorithmic}
    \caption{Private Memory Trace Generation}
    \label{alg:private_tr}
\end{algorithm}

The parallel sections of the OpenMP code are executed concurrently on different cores. Thus, each core has its copy of the parallel sections of the code. We mimic this behavior by making copies of each basic block's memory references under the parallel sections. Our mimicking strategy tries to replicate the memory trace of an OpenMP program on multiple cores. For example, suppose the parallel program is using $4$ cores. In that case, we make four copies of a basic block, we then add an {\em offset} to the memory addresses for each of the cores under execution except the core executing master thread. The offset is added to all memory references of a parallel region's basic blocks except for the shared variables' memory references. Some basic blocks (\emph{e.g.} loop iterations) under the parallel region are executed multiple times. So, they appear multiple times in the labeled memory trace. After adding offsets in the same way, we distribute the memory references belonging to these basic blocks evenly among all the cores. We choose the offset so that the mimicked memory references do not match the original memory references produced in the sequential execution. This mimicking strategy helps to show that the memory references belong to different cores.

The corresponding core accesses the private caches (such as $L_1$) to perform thread-specific execution. Therefore, we employ the procedure described in Algorithm~\ref{alg:private_tr} to generate private traces for each core. From the traces, we calculate corresponding reuse profiles and hit-rates for private caches. It takes a list of all the basic blocks, the number of cores, the sequential memory trace, the references belonging to shared variables, and basic block counts as input. The basic block counts denote the number of times each basic block is executed during program execution. If the count for a basic block is less than the number of cores, each thread gets a basic block copy. Otherwise, we distribute the count evenly among the core by setting the $bb\_count\_per\_core$ variable. In the next step, we traverse the memory trace and check whether it is a memory reference or not by checking the trace's entries. Suppose it is a memory reference and there is only one instance of the basic block in the memory trace it belongs to ($bb\_count\_per\_core[bb\_id]$ value is one). In that case, we make a copy of that reference for each core, add offset to the memory reference and assign it to each core. If the corresponding basic block has multiple instances in the memory trace ($bb\_count\_per\_core[bb\_id]$ value is greater than one), we assign it to a particular core. Note that if the reference belongs to the shared variables, then we do not add the offset. It is possible to perform the distribution with chunk size, similar to OpenMP static scheduling with chunk size. We find the list of basic blocks and corresponding counts using our LLVM based offline code analysis tool.

The OpenMP library can perform different scheduling strategies (\emph{e.g.}, static, dynamic, guided) while executing the code's parallel sections. Recording memory traces for each scheduling strategy is cumbersome and inefficient in terms of both time and memory. Thus, In this work, we generate a trace similar to the OpenMP scheduled traces. We use the recorded basic block labeled sequential trace to mimic the interleaving of threads. Our mimicking strategy distributes the corresponding memory trace equally among multiple threads under execution, similar to following static scheduling in OpenMP. To further study the effect of scheduling strategies on memory reuse, we propose various interleaving and scheduling strategies, described in section~\ref{sec:interleave}.

To generate the shared memory trace, we take the private traces and interleave the memory references. We use both \emph{round-robin} and \emph{uniform random} scheduling to interleave the memory references. Similar traces can be generated with binary instrumentation tools such as \emph{Valgrind}~\cite{valgrind}, \emph{memTrace}~\cite{payer2013lightweight} and \emph{Pin}~\cite{pin-tool,McCurdy-pin-interleave}. However, we use an LLVM based tool to leverage the conceptual advantage of dealing with simple straight line basic blocks within a program. \emph{Valgrind's Lackey} tool runs the multi-threaded program sequentially per thread, where the interleaving of the threads is left to the operating system. Therefore the resultant memory trace happens to be multi-threaded.

On the other hand, with \emph{Pin}, one has to collect the memory trace for a specific core count. Nonetheless, we cannot derive a basic block labeled trace from Pin instead of our LLVM instrumentation. We can generate both private and shared memory traces from basic block labeled sequential memory trace with our approach. Later we estimate the reuse profile for each trace once we have the memory traces that mimics the multicore execution.

\begin{algorithm}[t]
    \begin{algorithmic}[1]
    \Procedure{$interleave\_traces$}{$num\_cores$, $private\_mem\_traces$, $strategy$}
        \State $trace\_traverse\_done \gets [[]*num\_cores]$
        \State $interleaved\_trace \gets []$
        \For{$core$ $\textbf{in}$ $\textbf{range}(num\_cores)$}
            \State $trace\_traverse\_done[core] \gets false$
        \EndFor
        \State $core, i \gets 0$
        \While{$i < num\_cores$}
            \If{$strategy == uniform$} \label{line:startstrategy}
                \State $core \gets randint(0, num\_trces - 1)$ \textcolor{codegreen}{\Comment{Randomly choose core number}}
            \ElsIf{$strategy == round\_robin$}
                 \If{$core == num\_trces$}
                    \State $core \gets 0$
                \Else
                    \State $core \gets core + 1$ \textcolor{codegreen}{\Comment{Increase core number in round robin fashion}}
                \EndIf
            \EndIf \label{line:endstrategy}
            
            \If{$getline(line, private\_mem\_traces[core]) \neq -1$}
                \State $interleaved\_trace$ $\textbf{appends}$ $line$
            \Else
                \If{$trace\_traverse\_done[core] == false$}
                    \State $trace\_traverse\_done[core] \gets true$
                    \State $i \gets i + 1$
                \EndIf
            \EndIf
        \EndWhile
    \EndProcedure
    \end{algorithmic}
    \caption{Interleave memory traces}
    \label{alg:interleave_tr}
\end{algorithm}

\subsubsection{Interleaving Strategies}
\label{sec:interleave}

To determine the OpenMP application's shared cache performance, we employ multiple interleaving strategies to mimic the trace of shared memories. The shared trace is constructed as if all the threads in the application are executing concurrently and share the shared memory space (typically Last Level Cache). Algorithm \ref{alg:interleave_tr} show a high-level implementation of out interleaving strategies. It takes private traces, interleaving strategy, and the number of cores as inputs and applies our interleaving strategies to generate shared traces for shared memory accesses. We assume that the master thread is being executed in \emph{core 0}. We initiate an array \emph{trace\_traverse\_done} with boolean value false. The size of the array is the same as the number of cores under consideration. This is to track if all the memory references of the corresponding private trace have been traversed or not. Here, we employ two interleaving strategies: {\em round-robin} and {\em uniform-random} (see lines~\ref{line:startstrategy}--\ref{line:endstrategy}). In case of uniform random interleaving, we randomly choose any of the private traces and check if all the references from the trace have been read. If not, then we pick a reference from the trace and append it to the interleaved trace. For round-robin interleaving, we take the first memory reference from core 0's trace. Then we choose the next reference from the next core's trace. We repeat the process until all memory references from all traces are read and appended to interleaved trace. As our model is flexible, it is possible to implement other interleaving strategies in our model.

\vspace{-15pt}
\subsection{Cache Model}
We take the mimicked memory traces and cache configuration parameters as input and estimate the cache hit rates in this step. We use a reuse profile-based cache model to predict the cache hit rates. We further discuss the steps of our cache model as follows.

\vspace{-5pt}
\subsubsection{Reuse Profile Calculation}
We build the \textit{Private-stack} and the \textit{Concurrent} reuse profiles of the program (P(D)) from our mimicked private and shared memory traces. The conventional methods of measuring the reuse profile are costly because of the enormous size of the memory traces. The previous work of PPT used a stack-based method to compute the reuse profiles of a program~\cite{arafa_ics,arafa_ipccc,chennupati:pmbs}. The stack-based method has a worst-case time complexity of $\mathcal{O}(N.M)$ for a trace of length $N$ containing $M$ distinct references. In this work, we use a more optimized tree-based approach~\cite{PARDA:Niu} for calculating the reuse profiles from the memory trace, which has on average a time complexity of $\mathcal{O}(N.\log{}M)$.

\subsubsection{Hit Rate Estimation}
We measure private, and shared cache hit rates using an analytical memory model (SDCM) proposed by Brehob and Enbody~\cite{brehob:analytical}. SDCM was used before to predict the cache hit rates of CPUs~\cite{ppt-sasmm,chennupati:pmbs,chennupati:pads,ppt-amm} and GPUs~\cite{arafa_ics,arafa_ipccc} in PPT. Equation~\ref{eq:cond_phd} shows the way to measure the conditional hit rate at a given reuse distance ($P(h\mid D)$).
\begin{equation}\label{eq:cond_phd}
P(h\mid D) =  \sum_{a=0}^{A-1}\binom{D}{a}\biggl(\dfrac{A}{B}\biggr)^a\biggl(\dfrac{B-A}{B}\biggr)^{(D-a)}
\end{equation}
\noindent where {\em D} denotes the reuse distance, {\em A} denotes cache associativity and {\em B} denotes cache size in terms of number of blocks (which is cache size over cache line size). Typically, Eq.~\ref{eq:cond_phd} is used for an $n$-way associative cache. For a direct-mapped cache, conditional probability of hit is defined as

\begin{equation}\label{eq:phd_direct}
P\left(h\mid D\right) = \textit{$\biggl(\frac{B-1}{B}\biggr)^{D}$}
\end{equation}

Finally, we calculate the approximated unconditional probability of a hit {\em P(h)} for the entire program using Eq.~\ref{eq:phit}

\begin{equation}
\label{eq:phit}
P(h) = \sum\limits_{i=0}^N P({D_i}) \times P({h\mid D_i})
\end{equation}

\noindent where, $P({D_i})$ is the probability of $i^{th}$ reuse distance ($D$) in a reuse distribution $Pr(D)$. We further use these hit rates in the runtime prediction of the applications, which is beyond this paper's scope.

\vspace{-10pt}
\subsection{Runtime Prediction}
In our final step, we predict the runtime of the parallel application. In most cases, the benchmark kernels are within OpenMP parallel sections. When measuring parallel applications' performance, usually, we are interested in the parallel kernels and do not care about the sequential initialization and cleanup phases of the benchmarks. We take the ideas used by Chennupati \emph{et al.} in~\cite{chennupati:pmbs} to predict the runtimes of parallel sections of the benchmark applications. Chennupati \emph{et al.} predicted the runtimes of a sequential application where we predict the parallel application's runtimes.

Two main factors contribute to the runtime of a parallel application. Those are average time taken for the CPU operations (\textit{T$_{CPU}$}) and average memory access time (\textit{T$_{mem}$}). We use Byfl~\cite{Byfl}, an LLVM based application characterizing tool to get the number of CPU operations and the total memory required for the kernel execution. The parallel kernels' reuse profile is used along with the total memory required to determine memory latency.

Therefore, the predicted runtime is measured using Eq.~\ref{eq:total_runtime}

\begin{equation}\label{eq:total_runtime}
    T_{\text {pred}}=T_{mem}+T_{CPU}
\end{equation}

\subsubsection{Runtime Prediction considering Contiguous Memory Access}
If we assume that the available memory is contiguous, then the average memory access time is measured using equation~\ref{eq:cont-tmem}.

\begin{equation}\label{eq:cont-tmem}
T_{\text {cont-mem}}=\frac{\delta_{\text{avg}}+(b-1) \times \beta_{\text{avg}}}{b} \times total\_mem
\end{equation}

where $\delta_{\text {avg}}$, $\beta_{\text{avg}}$, \textit{b} and \textit{total\_mem} denote average latency, average reciprocal throughput, block size and the total memory (bytes) required by the program respectively. In the equation, we consider the average latency $\delta_{\text {avg}}$ and throughput $\beta_{\text{avg}}$ as per memory access, while we consider the block size \emph{b} as word size assuming the available memory is contiguous. By dividing $\delta_{\text{avg}}+(b-1) \times \beta_{\text {avg }}$ by block size \emph{b} we find average memory access time per byte. Multiplying this result with total\_mem provides total memory access time of a program in contiguous memory setting.

The average latency and throughput of a program depend on the hit rates at different cache levels. The average latency of a program on a machine with a three-level cache can be calculated using equation~\ref{eq:avg_latency}.

\begin{equation}\label{eq:avg_latency}
\begin{aligned}
    \delta_{\text {avg }}=P_{L_{1}}(h) \times \delta_{L_{1}} + \left(1-P_{L_{1}}(h)\right) \biggl[\biggr. P_{L_{2}}(h) \times \delta_{L_{2}}+\left(1-P_{L_{2}}(h)\right) \\
    \times \left[ P_{L_{3}}(h) \times \delta_{L_{3}} + \left(1-P_{L_{3}}(h)\right) \times \delta_{RAM} \right]\biggl. \biggr]
\end{aligned}
\end{equation}

where, $\delta_{L1}$, $\delta_{L2}$, $\delta_{L3}$, and $\delta_{RAM}$ are latencies of L1, L2, L3 caches and RAM respectively; $P_{L_{1}}(h), P_{L_{2}}(h) and P_{L_{3}}(h)$ are the probabilities of a hit for L1, L2 and L3 caches respectively.

Similarly, we measure the average throughput, $\beta_{\text {avg}}$ using equation~\ref{eq:avg_thrpt}.

\begin{equation}\label{eq:avg_thrpt}
\begin{aligned}
    \beta_{\text {avg }}=P_{L_{1}}(h) \times \beta_{L_{1}} + \left(1-P_{L_{1}}(h)\right) \biggl[\biggr. P_{L_{2}}(h) \times \beta_{L_{2}}+\left(1-P_{L_{2}}(h)\right) \\
    \times \left[ P_{L_{3}}(h) \times \beta_{L_{3}} + \left(1-P_{L_{3}}(h)\right) \times \beta_{RAM} \right]\biggl. \biggr]
\end{aligned}
\end{equation}

Using Byfl, we identify the number of CPU operations (ADD, SUB, DIV, etc.) in the parallel section. We divide those numbers by the number of cores and then measure \textit{T$_{CPU}$}, the time required for CPU operations for one core, using the hardware-specific instruction latencies and the operations count. We assume that the total workload is distributed among multiple cores evenly. Finally, the total runtime is predicted as $T_{pred}$.

\renewcommand{\arraystretch}{1.0}
\begin{table*}[t]
    \centering
    \caption{Benchmark applications used to verify our model. $\dagger$ and $\diamondsuit$ denote applications from PolyBench/OpenMP~\cite{polybench} and PARSEC~\cite{parsec} benchmark suites respectively. Last column shows the abbreviation of benchmark names which are used later to represent results.} 
    \begin{tabular}{|P{1.8cm}|P{2.4cm}|P{1.5cm}|P{1.7cm}|P{0.95cm}|P{0.9cm}|}
    \hline
    \textbf{Application} & \textbf{Description} & \textbf{Domain}  & \textbf{Input Size} & \textbf{Trace Size} & \textbf{Abbr.}\\
    \hline
    \hline
    ADI$^\dagger$ & Alternating Direction Implicit method for 2D heat diffusion & Stencils & 1k x 50 & 170GB & \textbf{adi}\\
    \hline
    ATAX$^\dagger$ & Matrix transpose and vector multiplication  & Linear Algebra & 4k x 4k & 14GB & \textbf{atx}\\
    \hline
    BICG$^\dagger$ & BiCG sub kernel of BiCGStab linear solver  & Linear Algebra & 4k x 4k & 11GB & \textbf{bcg}\\
    \hline
    Blackscholes$^\diamondsuit$ & Black-Scholes partial differential equation & Recognition, Mining and Synthesis & Options=64k, Runs=100 & 25GB & \textbf{blk}\\
    \hline
    Convolution-2D$^\dagger$ & 2D Convolution & Stencils & 4k x 4k & 19GB & \textbf{c2d}\\
    \hline
    Covariance$^\dagger$ & Covariance computation & Datamining & 1k x 1k & 208GB & \textbf{cov}\\
    \hline
    Doitgen$^\dagger$ & Multiresolution analysis kernel & Linear Algebra & 128x128x128 & 146GB & \textbf{dgn}\\
    \hline
    Durbin$^\dagger$ & Toeplitz system solver & Linear Algebra & 4k & 7.7GB & \textbf{dbn}\\
    \hline
    Gramschmidt$^\dagger$ & QR decomposition with modified Gram Schmidt & Linear Algebra & 512 x 512 & 64GB & \textbf{grm}\\
    \hline
    Jacobi$^\dagger$ & Jacobi iteration & Stencils & 4k x 4k & 41GB & \textbf{jcb}\\
    \hline
    LU$^\dagger$ & LU decomposition without pivoting & Linear Algebra & 1k & 179GB & \textbf{lu}\\
    \hline
    2MM$^\dagger$ & Two matrix multiplication & Linear Algebra & 512 x 512 & 115GB & \textbf{2mm} \\
    \hline
    MVT$^\dagger$ & Matrix vector product and transpose & Linear Algebra & 4k & 14GB & \textbf{mvt} \\
    \hline
    SYMM$^\dagger$ & Symmetric matrix-multiply & Linear Algebra & 1024 x 1024 & 335GB & \textbf{smm} \\
    \hline
    \end{tabular}
    \label{table:benchmarks}
\vspace{-17pt}
\end{table*}

\subsubsection{Runtime Prediction considering Non-contiguous Memory Access}

In reality, the memory alignment is non-contiguous. Therefore, there will be gaps $(\upsilon)$ in between the required program data. As a result, the new block size \emph{$b^{new}=b+\upsilon$} is used in equation~\ref{eq:cont-tmem} for determining non-contiguous memory access time. If the block size is large, the entire block may not be transferred from main memory to caches due to limiting factors such as cache size, data bus width, etc. Therefore, we model such memory access behavior as follows. Considering $b^{new}_{1}, b^{new}_{2}, b^{new}_{3}, ..., b^{new}_{i}, ..., b^{new}_{n}$ are the blocks of data on main memory and C be the amount of data transferred to cache from main memory at any given time, the new block size $b^{new}$ at a given cache size (B) can be re-written as:

\[\label{eq:bnew}
b^{new}= 
\begin{cases}
C              & :\textbf{if}\;\;\; b^{new}_i \leq C\\
\Bigg\lceil\dfrac{b^{new}_i}{C}\Bigg\rceil \times C & :\textbf{if}\;\;\; S \geq b^{new}_i \geq C\\
S              & :\textbf{if}\;\;\; b^{new}_i \geq S\\
\end{cases}
\]

In case of the time taken for CPU operations, there is a large difference in the instruction latencies between DIV and the rest of the instructions. Moreover, the time required for CPU operations is dependent on program characteristics, where some applications are instruction latency dependent while others are throughput reliable. Thus, the time for the resultant CPU operations (\textit{T$_{CPU}$}) can be written as:

\[\label{eq:TCPU}
T_{CPU}= 
\begin{cases}
\delta_{in} + (N_{in}-N_{in\_div}-1) \times \beta_{in}+\\
\delta_{div} + (N_{in\_div}-1) \times \beta_{div} & :\textit{throughput}\\ 
\\
(N_{in}-N_{in\_div}-1) \times \delta_{in} + \\
(N_{in\_div}-1) \times \delta_{div} & :\textit{latency}\\
\end{cases}
\]

where $\delta_{in}, \delta_{div}, \beta_{in}, \beta_{div}$ are latencies and throughputs of instructions, ADD/SUB, MUL and DIV respectively, while $N_{in}$ and $N_{in\_div}$ are the number of instructions.

\renewcommand{\arraystretch}{1.0}
\begin{table}[b]
    \centering
    \caption{Target CPUs}
    \label{table:cpus_for_veri}
    \begin{tabular}{|P{0.14\linewidth}|P{0.14\linewidth}|P{0.08\linewidth}|P{0.10\linewidth}|c|c|c|}
        \hline
        \textbf{Processor} & \textbf{Microarch-itecture} & \textbf{Core Count} & \textbf{Freq.} &  \textbf{L1} & \textbf{L2} & \textbf{L3} \\
        \hline
        Intel Core i7-5960X & Haswell & 8 & 3.0 GHz & 32 KB & 256 KB & 20 MB \\
        \hline
        Intel Xeon E5-2699 v4 & Broadwell & 22 & 2.2 GHz & 32 KB & 256 KB & 55 MB \\
        \hline
        AMD EPYC 7702P & Zen 2 & 64 & 2.0 GHz & 2 MB & 32 MB & 256 MB \\
        \hline
    \end{tabular}
\end{table}

\renewcommand{\arraystretch}{1.5}
\begin{table}[b]
    \centering
    \caption{PAPI events to measure cache hit rates on real machine}
    \label{tab:papi_events}
    \begin{tabular}{|c|c|}
    \hline
        \textbf{Event} & \textbf{PAPI events and equation used}\\
        \hline
        L1 D-Cache Hit Rate & $\text{1.0 - (PAPI\_L1\_DCM / (PAPI\_LD\_INS + PAPI\_SR\_INS)}$) \\
        \hline
        L2 Cache Hit Rate & $\text{1.0 - (PAPI\_L2\_DCM / (PAPI\_LD\_INS + PAPI\_SR\_INS)}$) \\
        \hline
        L3 Cache Hit Rate & $\text{1.0 - (PAPI\_L3\_TCM / (PAPI\_LD\_INS + PAPI\_SR\_INS)}$) \\
        \hline
    \end{tabular}
\end{table}

\section{Experiments and  Results}
\label{sec:results}

In this section, we validate our model and present the results. Table~\ref{table:benchmarks} shows a list of the applications used for validation. We use various applications representing different domains from PolyBench~\cite{polybench} and PARSEC~\cite{parsec} benchmark suites. For PolyBench, we use the OpenMP implementation by Grauer-Gray \emph{et al.}~\cite{polybench-acc}. We choose these benchmark suites as they are widely used for validating performance models. We use the standard input sizes for all the benchmarks. The generated basic block labeled memory trace sizes are also shown for the input used for each application.

Our model validations are two-fold: 1) hit-rates and 2) runtimes. For the validations, we model three processor architectures listed in table~\ref{table:cpus_for_veri}. We disable both hardware prefetching and hyperthreading for these machines' architectures as we do not model those features in this work. We bind a single thread to run on a single core. For the experiments, we start the core count with $1$ and increase it with a power of $2$; therefore, we report hit rates and execution times for \{1, 2, 4, 8\} cores for all the benchmarks. For Xeon and EPYC processors, we also show the results for the $16$ core configuration. We present the validation results of hit-rates in section~\ref{sec:Hit_Rate_Veri} and runtimes in section~\ref{sec:runtime_veri}.

\begin{figure*}[htbp]
    \centering
    \subfigure[{Cache hit rates when applications run on 1-core configuration}]{
    \label{fig:i7-hitrate-1Core}%
    \includegraphics[width=0.98\linewidth,height=4.1cm]{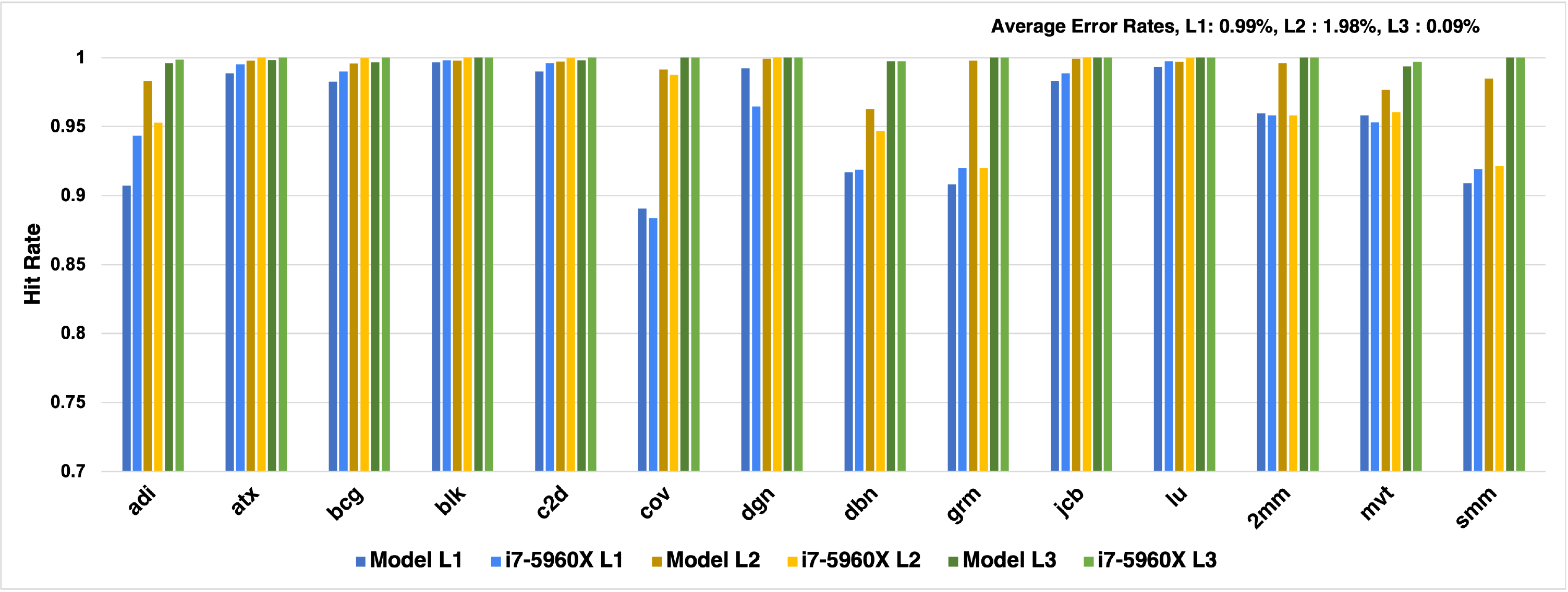}}%
    \qquad
    \subfigure[{Cache hit rates when applications run on 2-core configuration}]{
    \label{fig:i7-hitrate-2Core}%
    \includegraphics[width=0.98\linewidth,height=4.1cm]{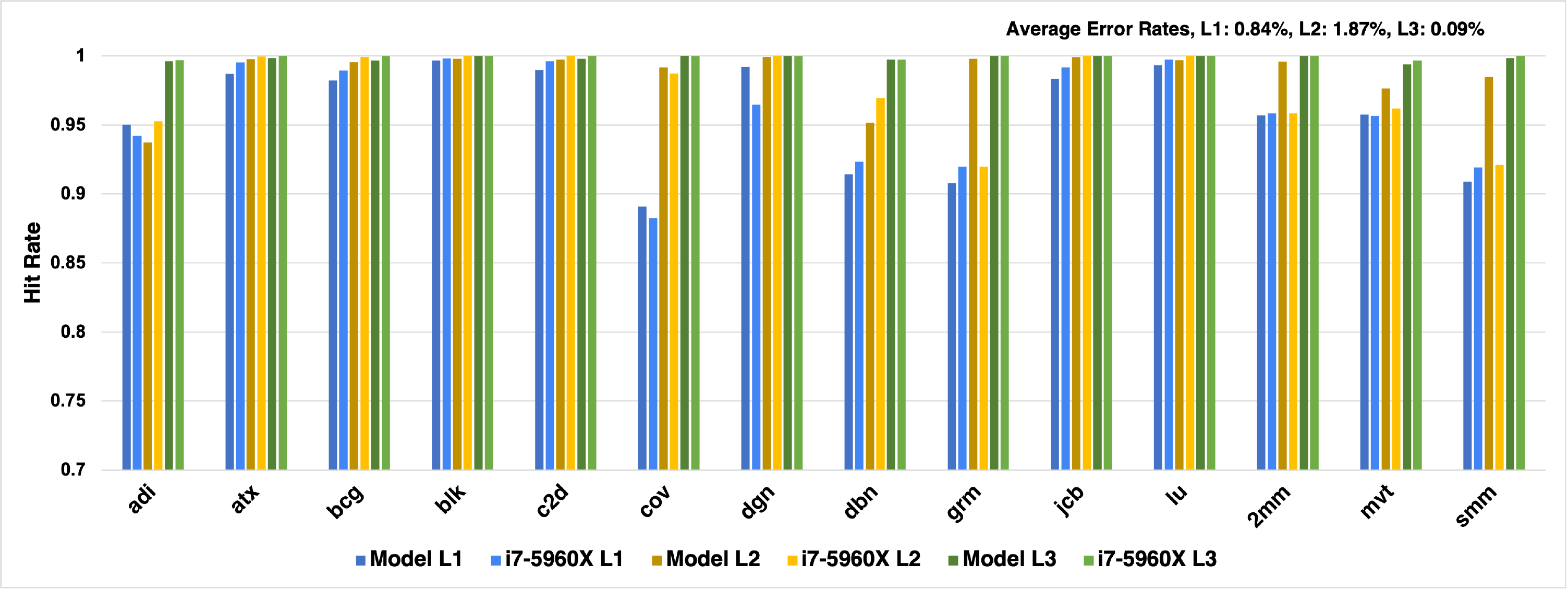}}%
    \qquad
    \subfigure[{Cache hit rates when applications run on 4-core configuration}]{
    \label{fig:i7-hitrate-4Core}%
    \includegraphics[width=0.98\linewidth,height=4.1cm]{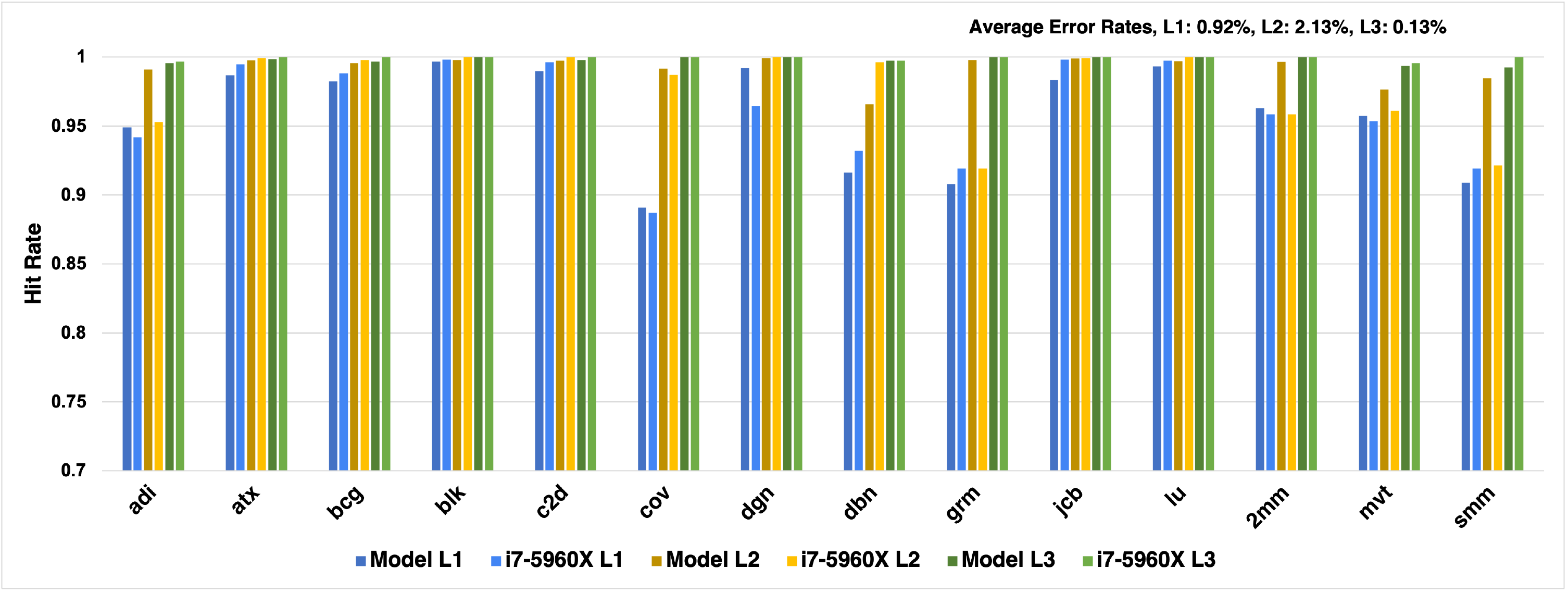}}%
    \qquad
    \subfigure[{Cache hit rates when applications run on 8-core configuration}]{
    \label{fig:i7-hitrate-8Core}%
    \includegraphics[width=0.98\linewidth,height=4.1cm]{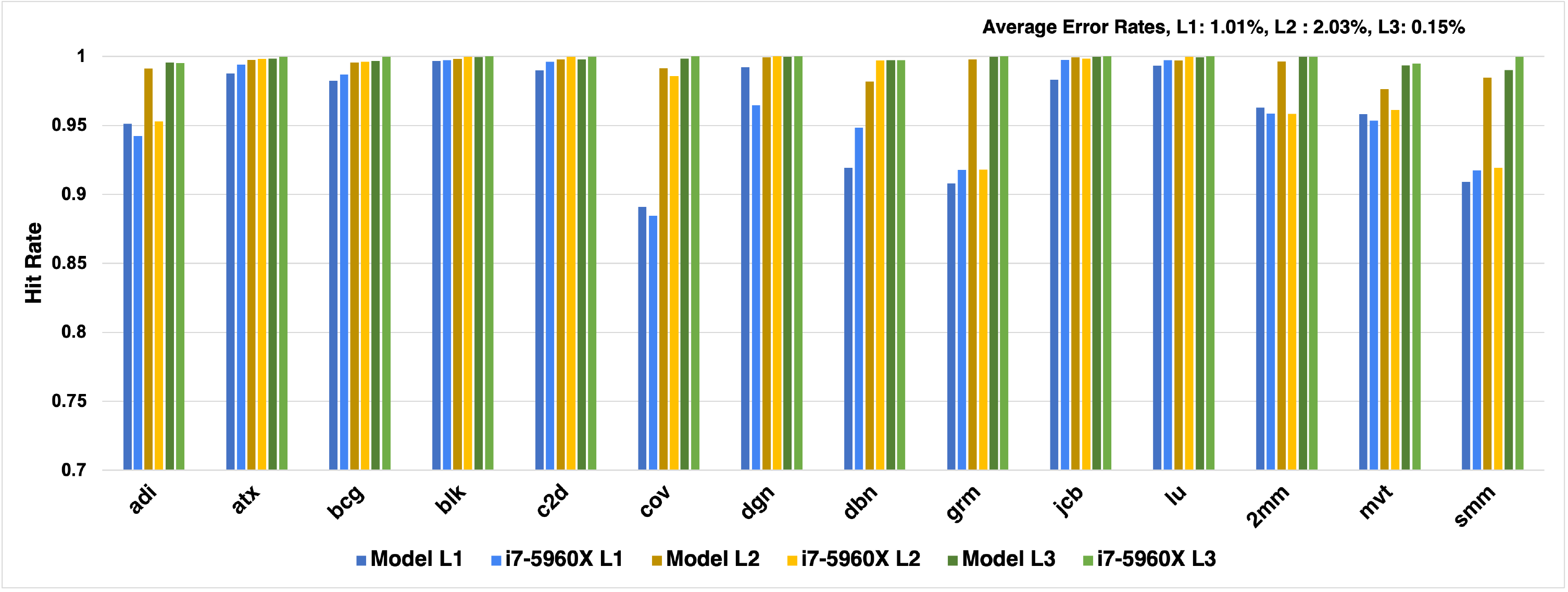}}%
    \qquad
    \caption{Hit rate comparison on Intel Core I7-5960X caches for different core configurations. Note that we start the Y-axis at $0.7$ on order to zoom in on the difference in performance.}
    \label{fig:results-hitrates-i7}%
\end{figure*}

\begin{figure*}[htbp]
    \centering
    \subfigure[{Cache hit rates when applications run on 1-core configuration}]{
    \label{fig:xeon-hitrate-1Core}%
    \includegraphics[width=0.98\linewidth,height=3.2cm]{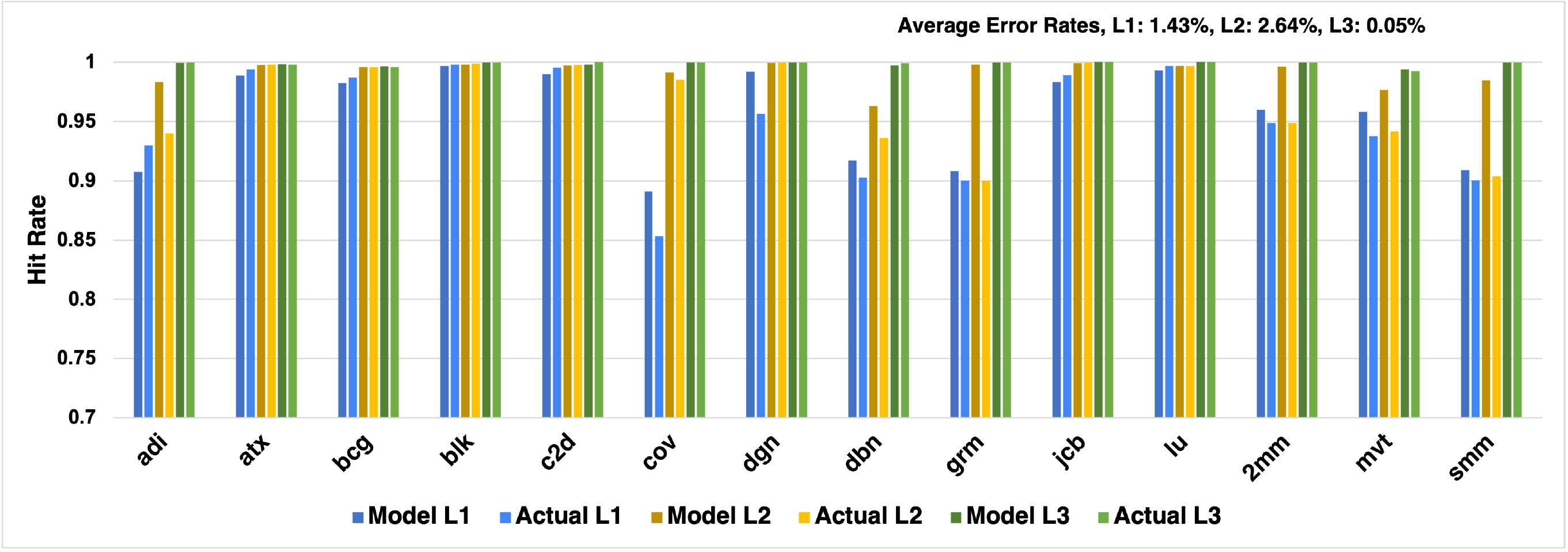}}%
    \qquad
    \subfigure[{Cache hit rates when applications run on 2-core configuration}]{
    \label{fig:xeon-hitrate-2Core}%
    \includegraphics[width=0.98\linewidth,height=3.15cm]{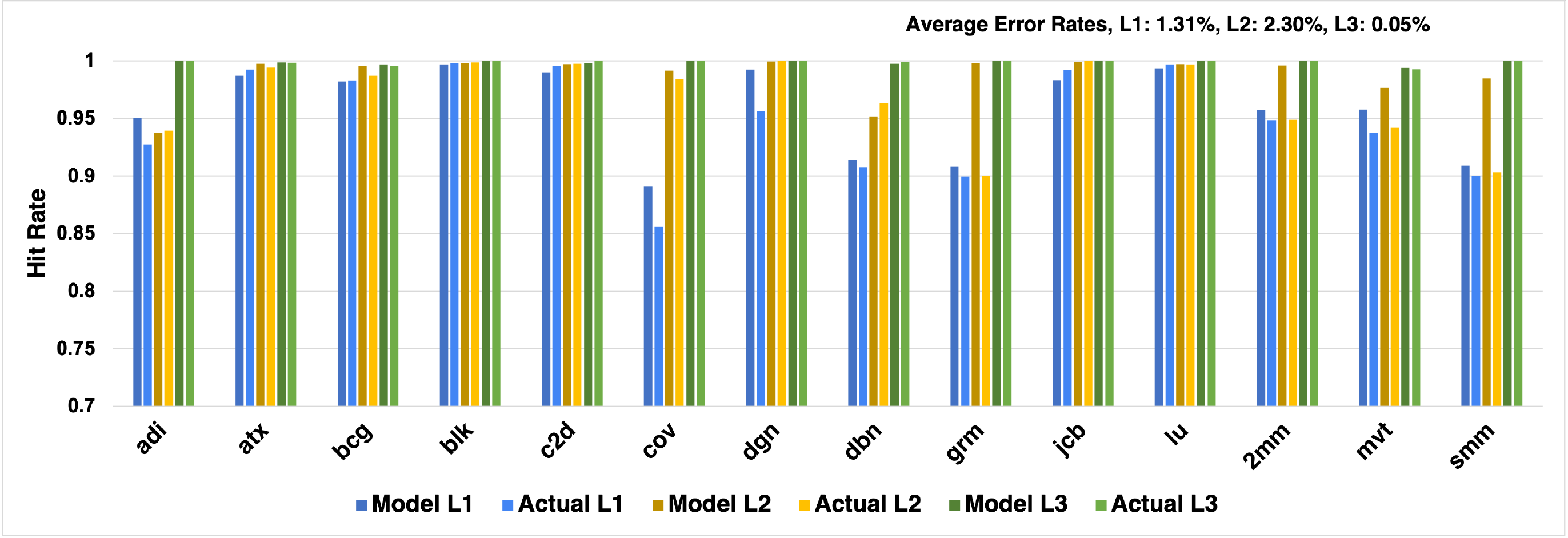}}%
    \qquad
    \subfigure[{Cache hit rates when applications run on 4-core configuration}]{
    \label{fig:xeon-hitrate-4Core}%
    \includegraphics[width=0.98\linewidth,height=3.15cm]{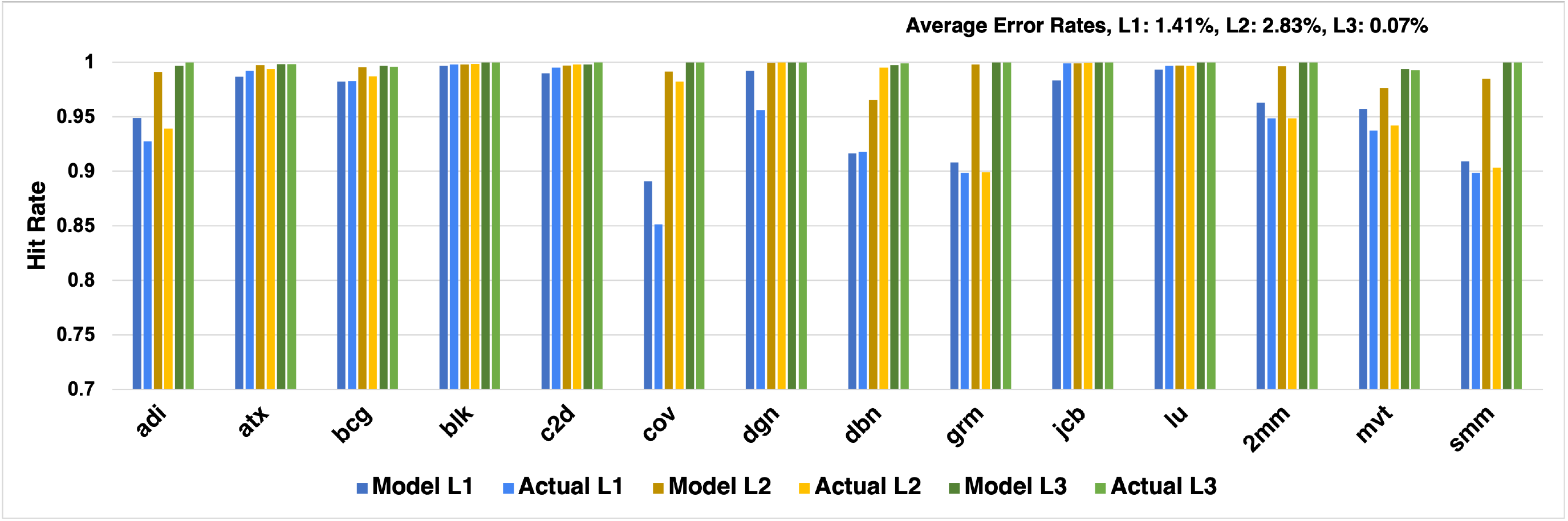}}%
    \qquad
    \subfigure[{Cache hit rates when applications run on 8-core configuration}]{
    \label{fig:xeon-hitrate-8Core}%
    \includegraphics[width=0.98\linewidth,height=3.15cm]{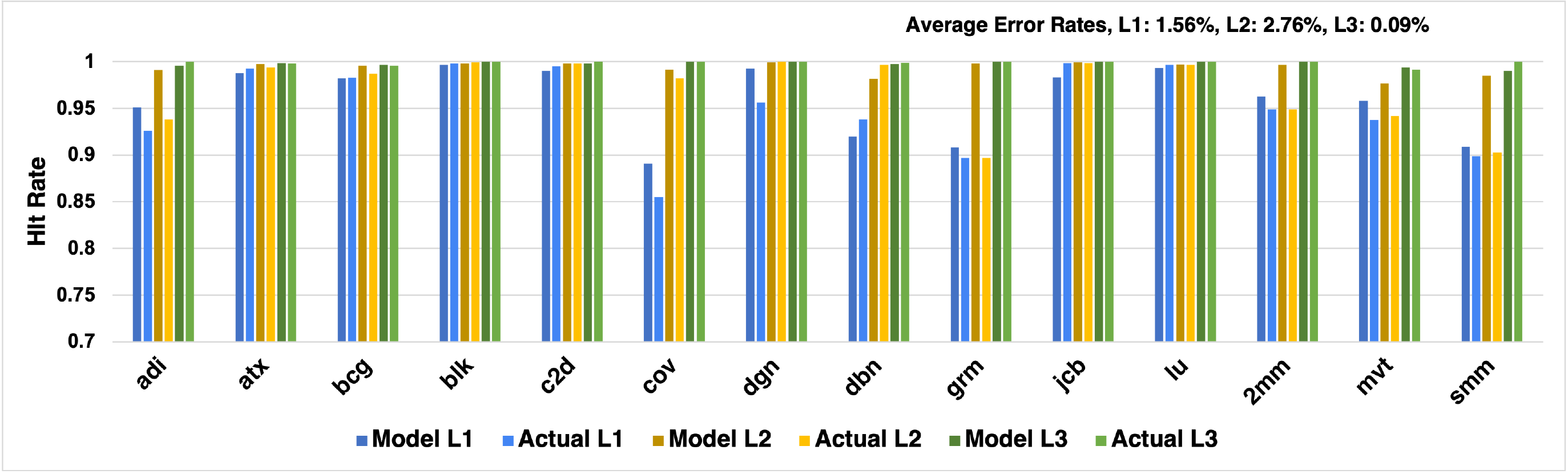}}%
    \qquad
    \subfigure[{Cache hit rates when applications run on 16-core configuration}]{
    \label{fig:xeon-hitrate-16Core}%
    \includegraphics[width=0.98\linewidth,height=3.15cm]{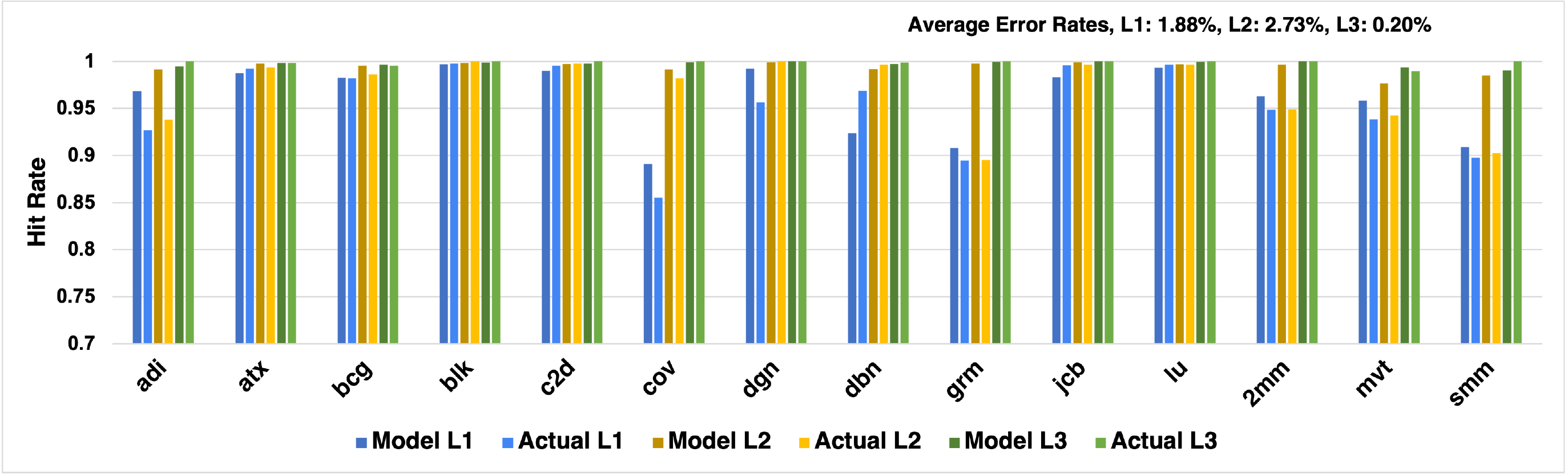}}%
    \qquad
    \caption{Hit rate comparison on Intel Xeon E5-2699 caches for different core configurations. Note that we start the Y-axis at $0.7$ on order to zoom in on the difference in performance.}
    \label{fig:results-hitrates-xeon}%
\end{figure*}

\vspace{-14pt}
\subsection{Cache Hit Rate Verification}
\label{sec:Hit_Rate_Veri}
\vspace{-5pt}
We validate the predicted hit-rates from our cache model against the PAPI~\cite{papi-c} performance counters. Table~\ref{tab:papi_events} shows the PAPI events (equations) used to determine the cache hit rates on real hardware. We report hit-rates for varying numbers of cores \{1, 2, 4, 8\} for all the benchmarks across all the experimental architectures. While collecting the PAPI performance counters to calculate hit rates, we change the number of threads/cores using \textit{$OMP\_NUM\_THREADS$} environment variable.

Most of the benchmark applications that we use have serial codes to initialize the data structures that are later accessed in parallel. Thus, the data that we access in parallel sections might already be in the cache due to their previous access to the applications' initialization. This makes the cache warm cache (requested data is already in the cache). However, in our cache model, we assume the caches are cold (cache is empty when the first data is requested). To get around this issue, we add a large dummy array access to the code before starting the PAPI counters but after the program's initialization. This way, we evict all the data loaded into the cache due to the initialization section's execution. As a result, when the parallel kernels are executed, the cache acts as a cold cache.

Figure~\ref{fig:results-hitrates-i7} compares the hit rates on Intel Core i7-5960X for each of the core configurations \{1, 2, 4, and 8 cores\}. We show the hit rates of the benchmarks for Intel Core i7-5960X in Figures~\ref{fig:i7-hitrate-1Core}--\ref{fig:i7-hitrate-8Core} respectively. For L1 cache our model’s average error rates are 0.99\%, 0.84\%, 0.92\%, and 1.01\% for the core configurations where 1.98\%, 1.87\%, 2.13\%, and 2.03\% are average error rates for L2 and 0.09\%, 0.09\%, 0.13\%, and 0.15\% are the average error rates for L3. The results for Xeon E5-2699 processor are shown in figure~\ref{fig:results-hitrates-xeon} for 1, 2, 4, 8 and, 16 core configurations. For Xeon's L1 cache, our model’s average error rates are 1.43\%, 1.31\%, 1.41\%, 1.56\%, and 1.88\% for the respective core configurations where 2.64\%, 2.30\%, 2.83\%, 2.76\% and 2.73\% are average error rates for L2 and 0.05\%, 0.05\%, 0.07\%, 0.09\% and 0.20\% are the average error rates for L3. The results show that our model predicts the hit rates of caches accurately with an overall average error rate of \textbf{1.23\%}.

The PAPI events listed in table~\ref{tab:papi_events} are not available on the AMD EPYC 7702P processor. Therefore, we can not verify the predicted cache hit rates for this CPU with real hardware.

As we mimic the memory traces of multi-threaded execution on different cache levels from single-threaded execution, we do not consider the effect of cache coherence in our model. Still, the experiments show promising results. Overall, our error rates for cache hit rate prediction are reasonable. However, we notice a slightly higher error rate for \emph{Gramschmidt} and \emph{SYMM} when predicting L2 cache hit rates on both processors irrespective of core count. On Core i7-5960X, the average error for \emph{Gramschmidt's} L2 hit rate prediction is 8.55\%, where on Xeon E5-2699, it is 11.09\%. For \emph{SYMM} the error average rates are 6.94\% and 9.04\%.

\begin{figure}[b]
    \centering
    \lstset{style=mystyle}
    \begin{lstlisting}[language=python,numbers=left]
tasklist = [['iALU', n_iALU/num_cores], ['fALU', n_fALU/num_coress], ['fDIV', n_fDIV/num_cores], ['MEM_ACCESS', reuse_dist_shared, probability_rd_shared, block_size, mem_in_bytes, data_bus_width, num_cores, reuse_dist_private_list, probability_rd_private_list]]

time = core.time_compute_multicore(tasklist)
    \end{lstlisting}
    \caption{An example of tasklist passed to PPT}
    \label{fig:tasklist}
\end{figure}

\begin{figure}[tb]
\centering
\subfigure[{Applications running on 1 core}]{
\label{fig:i7-runtime-1Core}%
\includegraphics[width=0.46\linewidth]{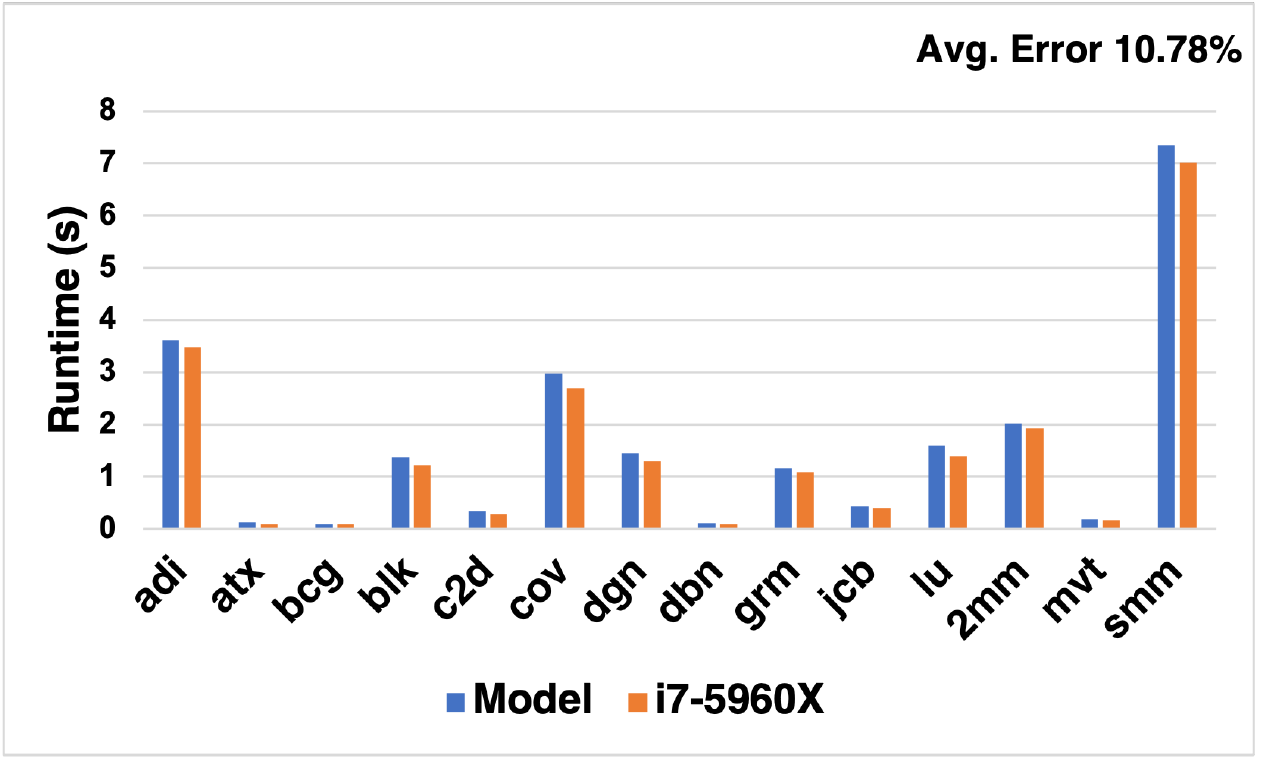}}
\qquad
\subfigure[{Applications running on 2 cores}]{
\label{fig:i7-runtime-2Core}%
\includegraphics[width=0.46\linewidth]{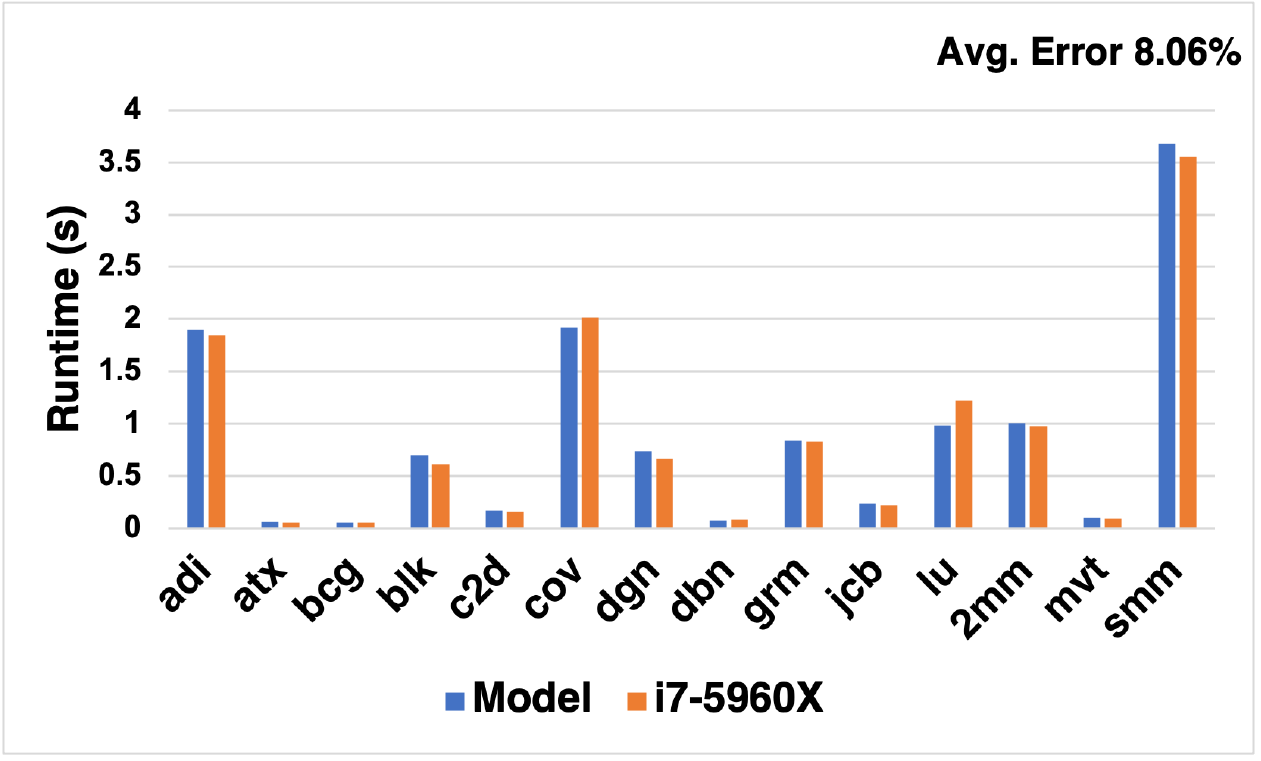}}
\qquad
\subfigure[{Applications running on 4 cores}]{
\label{fig:i7-runtime-4Core}%
\includegraphics[width=0.46\linewidth]{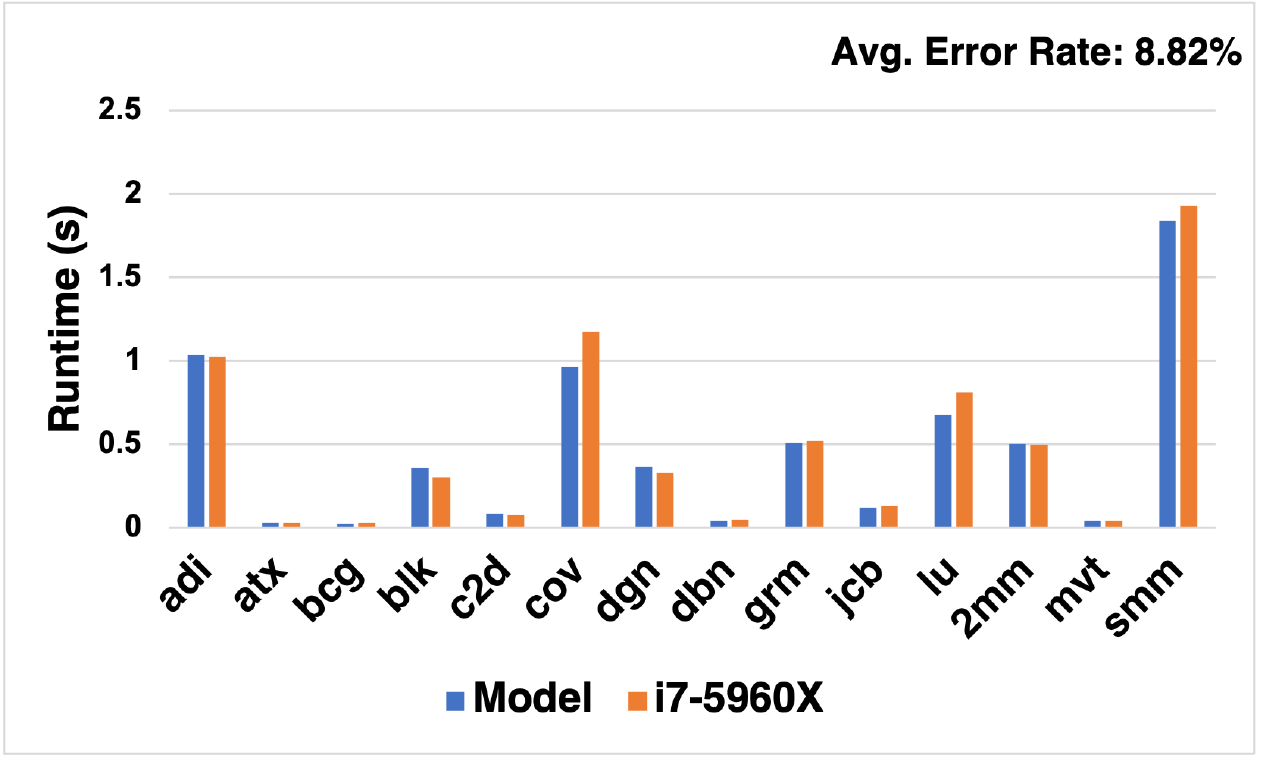}}
\qquad
\subfigure[{Applications running on 8 cores}]{
\label{fig:i7-runtime-8Core}%
\includegraphics[width=0.46\linewidth]{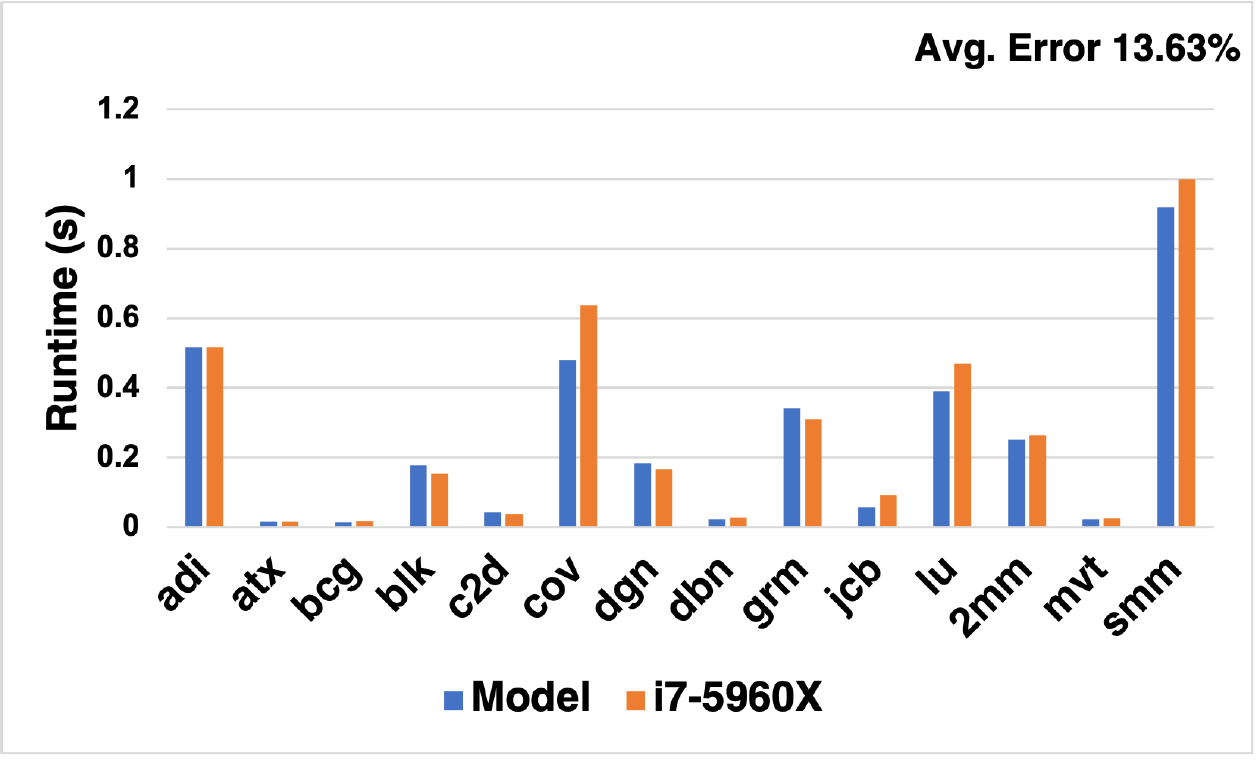}}
\qquad
\caption{OpenMP applications' parallel sections' runtime comparison on Intel Core I7-5960X with different core configurations}
\label{fig:results-runtime-i7}%
\end{figure}

\begin{figure}[htbp]
\centering
\subfigure[{Applications running on 1 core}]{
\label{fig:xeon-runtime-1Core}%
\includegraphics[width=0.46\linewidth]{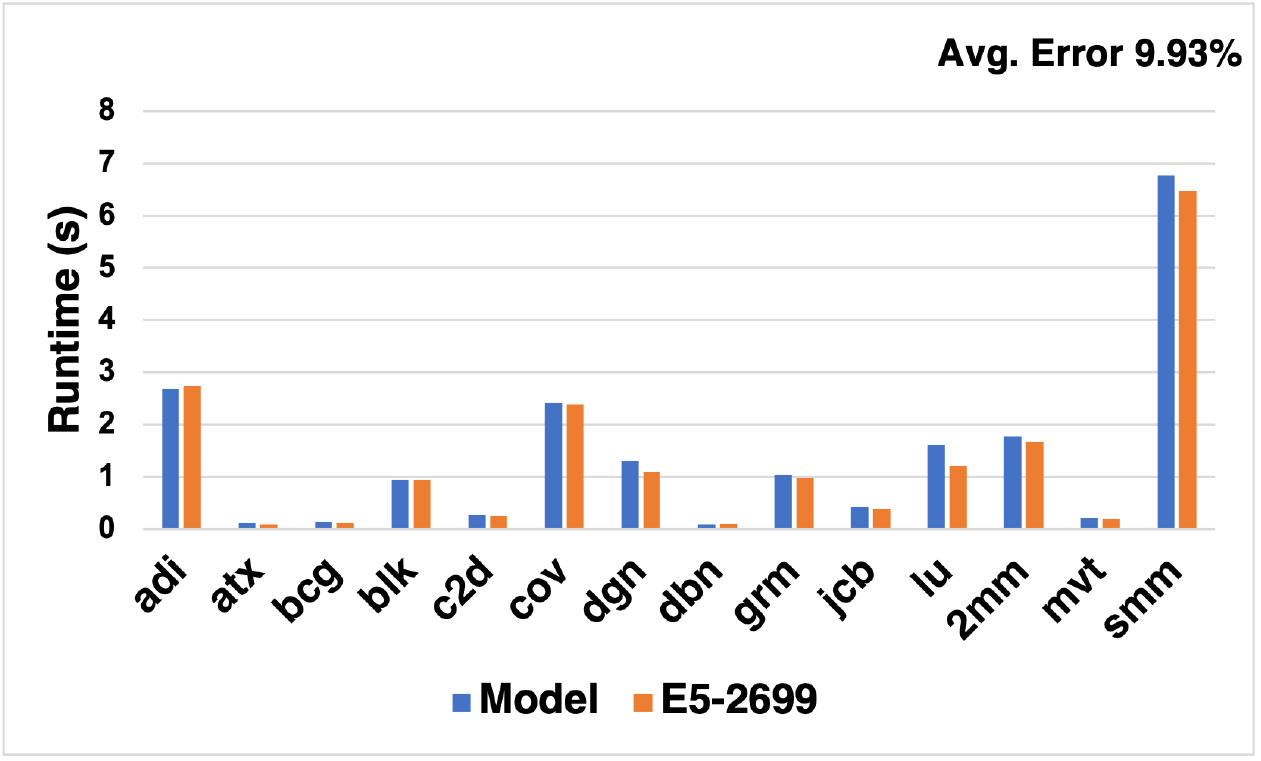}}%
\vspace{3pt}
\qquad
\subfigure[{Applications running on 2 cores}]{
\label{fig:xeon-runtime-2Core}%
\includegraphics[width=0.46\linewidth]{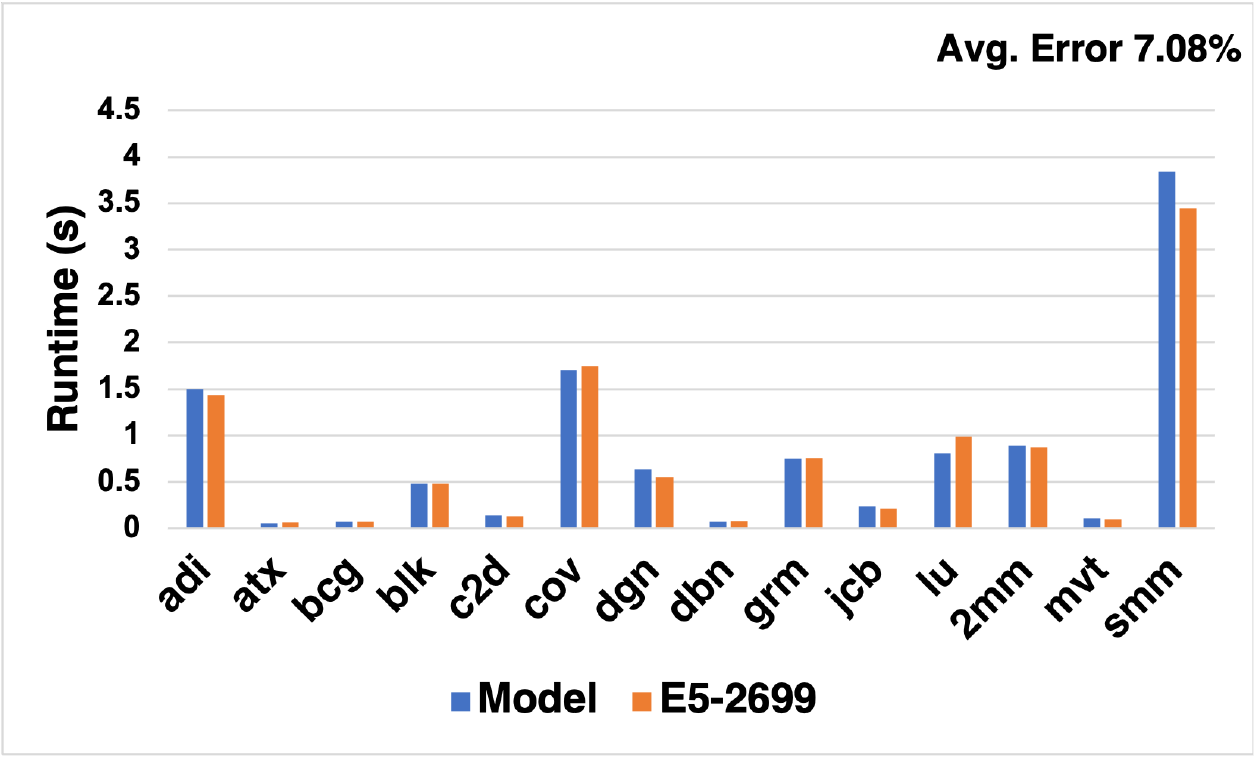}}%
\vspace{3pt}
\qquad
\subfigure[{Applications running on 4 cores}]{
\label{fig:xeon-runtime-4Core}%
\includegraphics[width=0.46\linewidth]{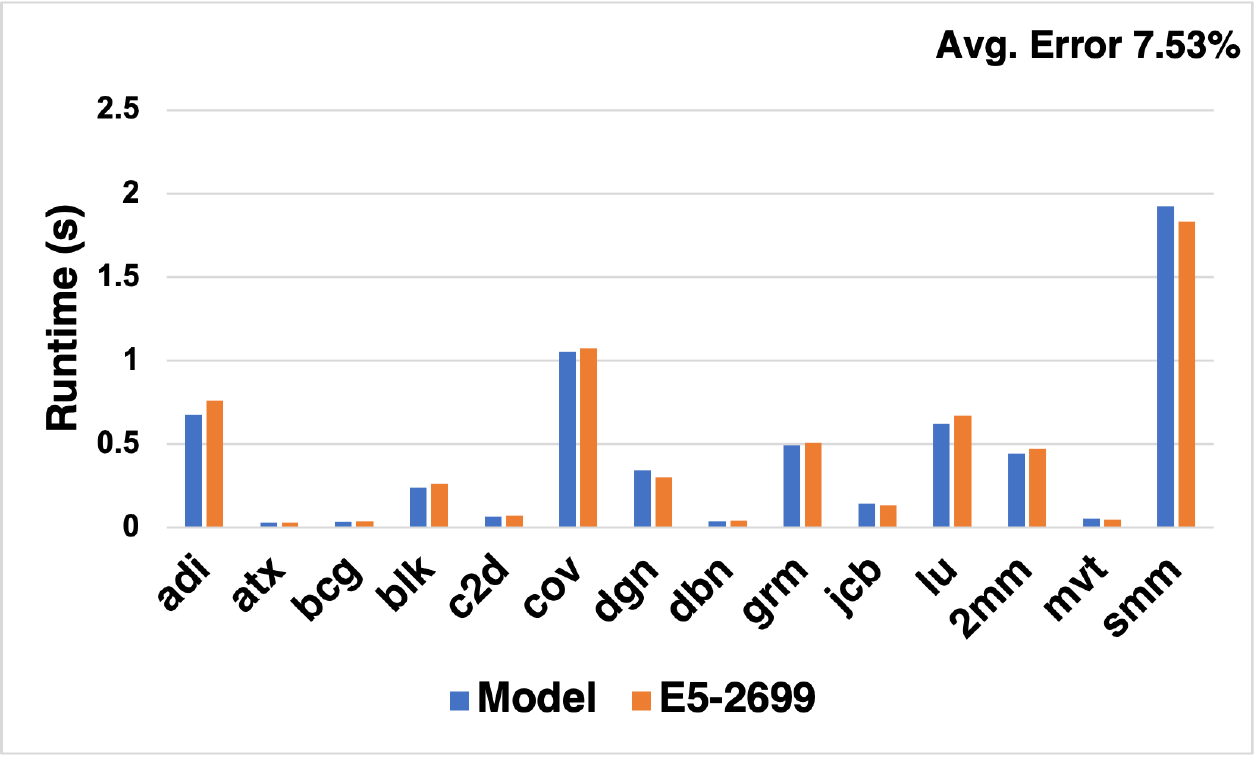}}%
\vspace{3pt}
\qquad
\subfigure[{Applications running on 8 cores}]{
\label{fig:xeon-runtime-8Core}%
\includegraphics[width=0.46\linewidth]{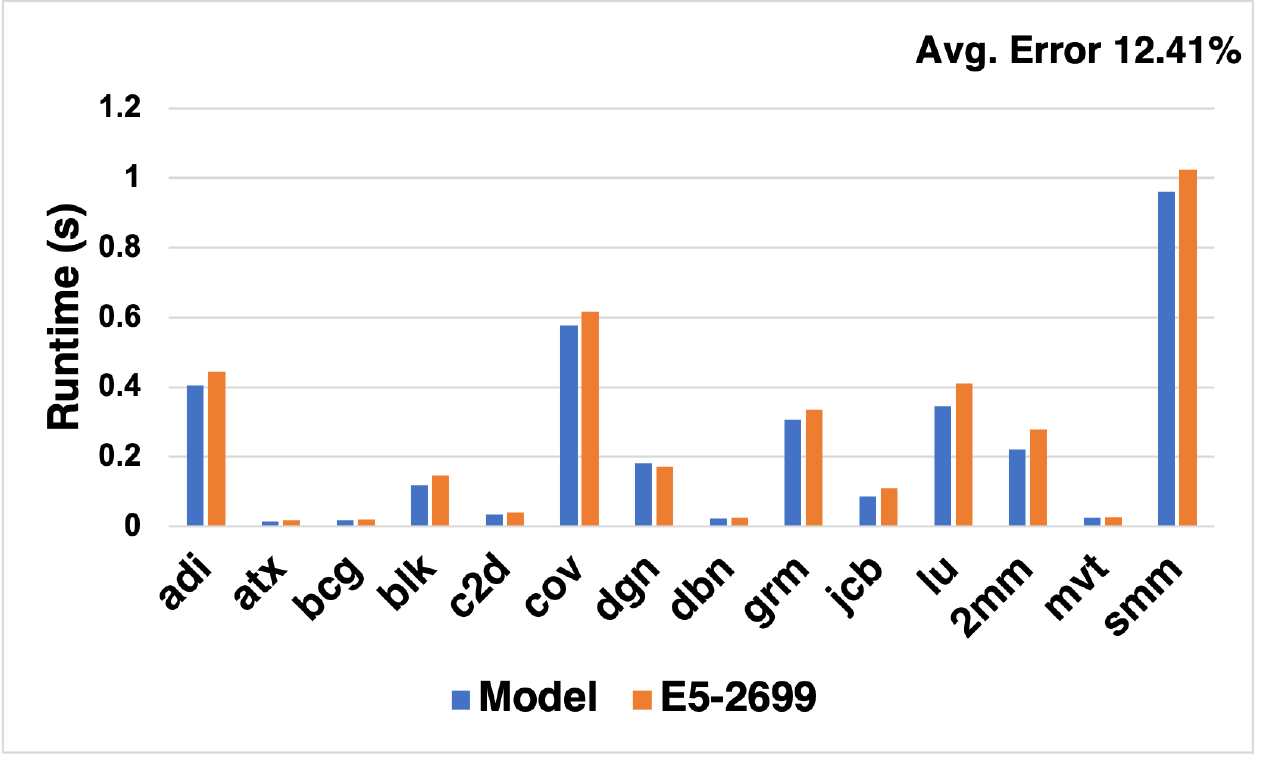}}%
\vspace{3pt}
\qquad
\subfigure[{Applications running on 16 cores}]{
\label{fig:xeon-runtime-16Core}%
\includegraphics[width=0.46\linewidth]{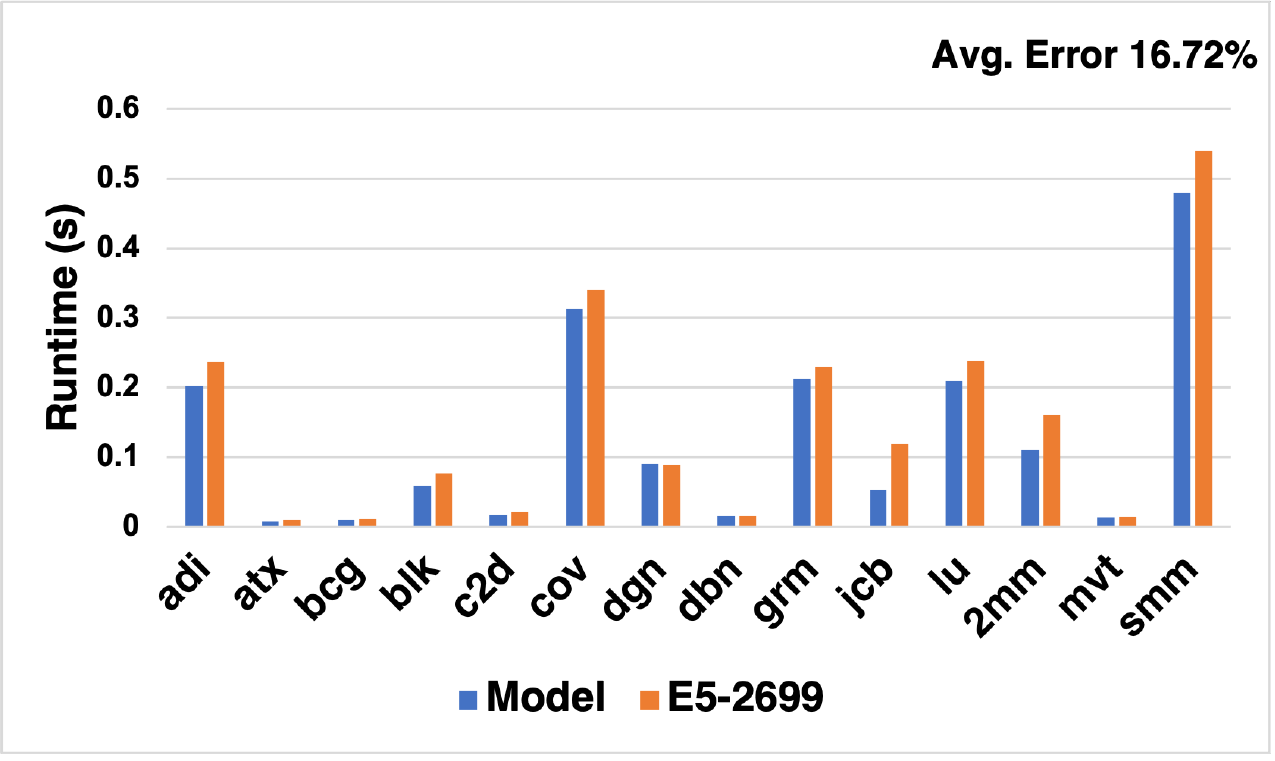}}%
\vspace{3pt}
\qquad
\caption{OpenMP applications' parallel sections' runtime comparison on Intel Xeon E5-2699 with different core configurations}
\label{fig:results-runtime-xeon}%
\end{figure}

\begin{figure}[htbp]
\centering
\subfigure[{Applications running on 1 core}]{
\label{fig:epyc-runtime-1Core}%
\includegraphics[width=0.46\linewidth]{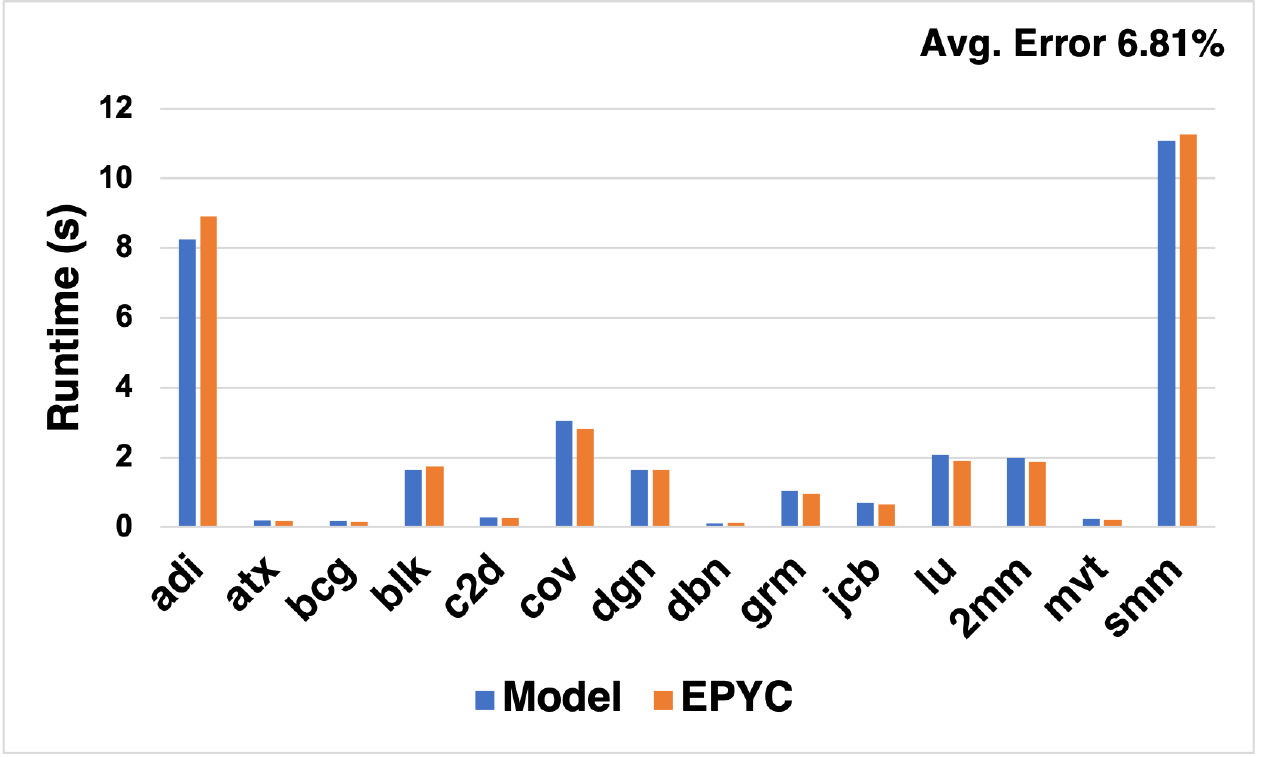}}%
\qquad
\subfigure[{Applications running on 2 cores}]{
\label{fig:epyc-runtime-2Core}%
\includegraphics[width=0.46\linewidth]{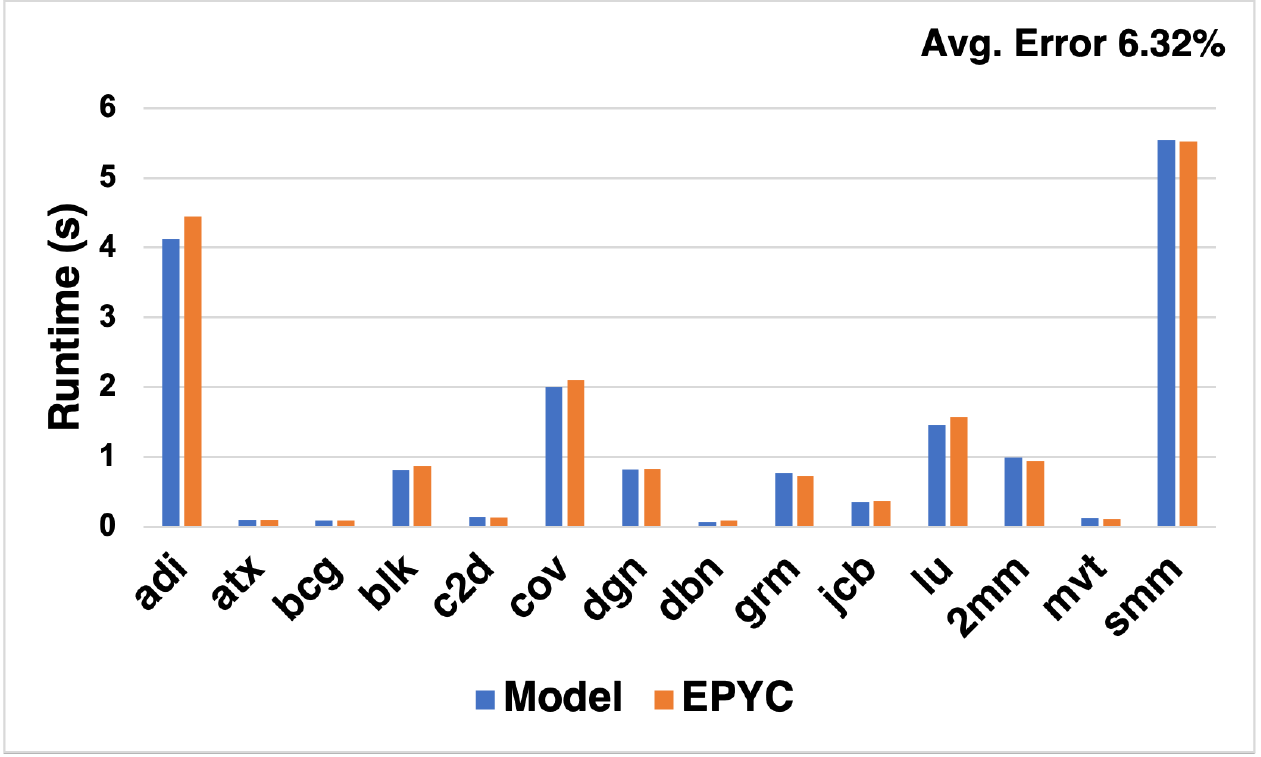}}%
\qquad
\subfigure[{Applications running on 4 cores}]{
\label{fig:epyc-runtime-4Core}%
\includegraphics[width=0.46\linewidth]{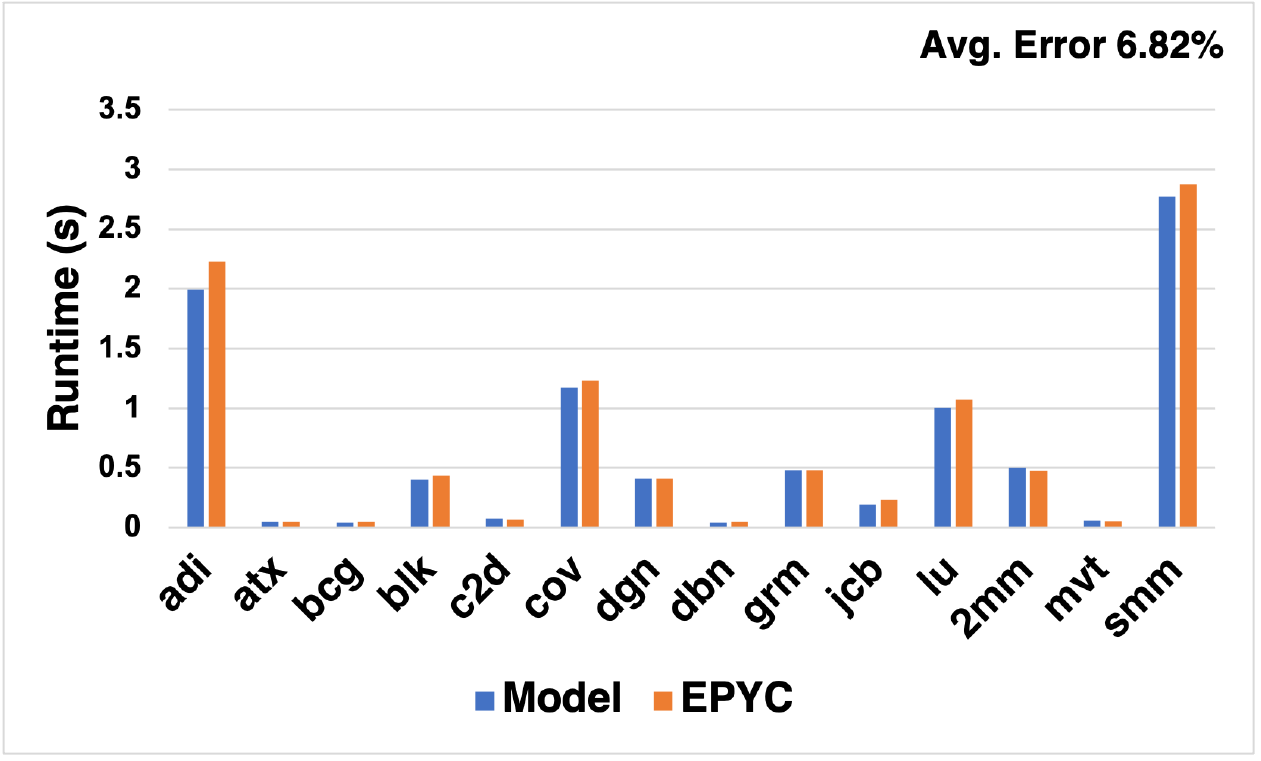}}%
\qquad
\subfigure[{Applications running on 8 cores}]{
\label{fig:epyc-runtime-8Core}%
\includegraphics[width=0.46\linewidth]{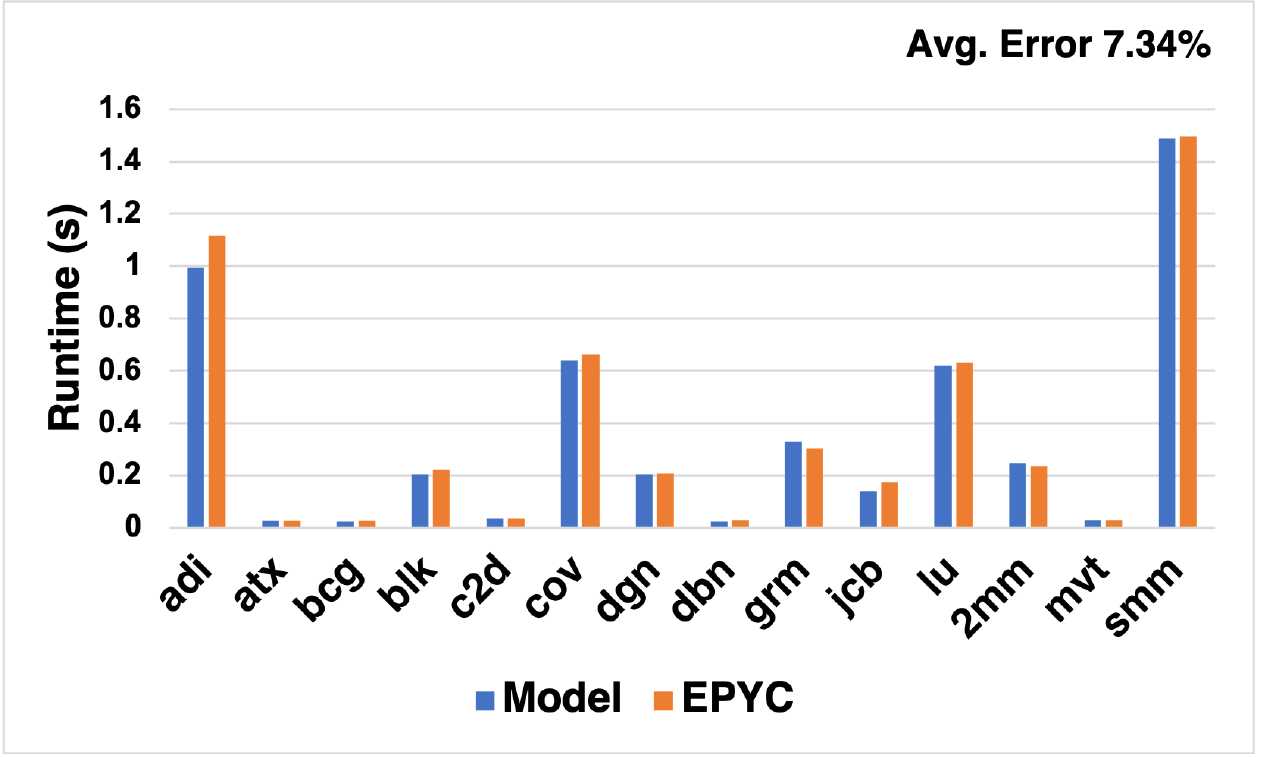}}%
\qquad
\subfigure[{Applications running on 16 cores}]{
\label{fig:epyc-runtime-16Core}%
\includegraphics[width=0.46\linewidth]{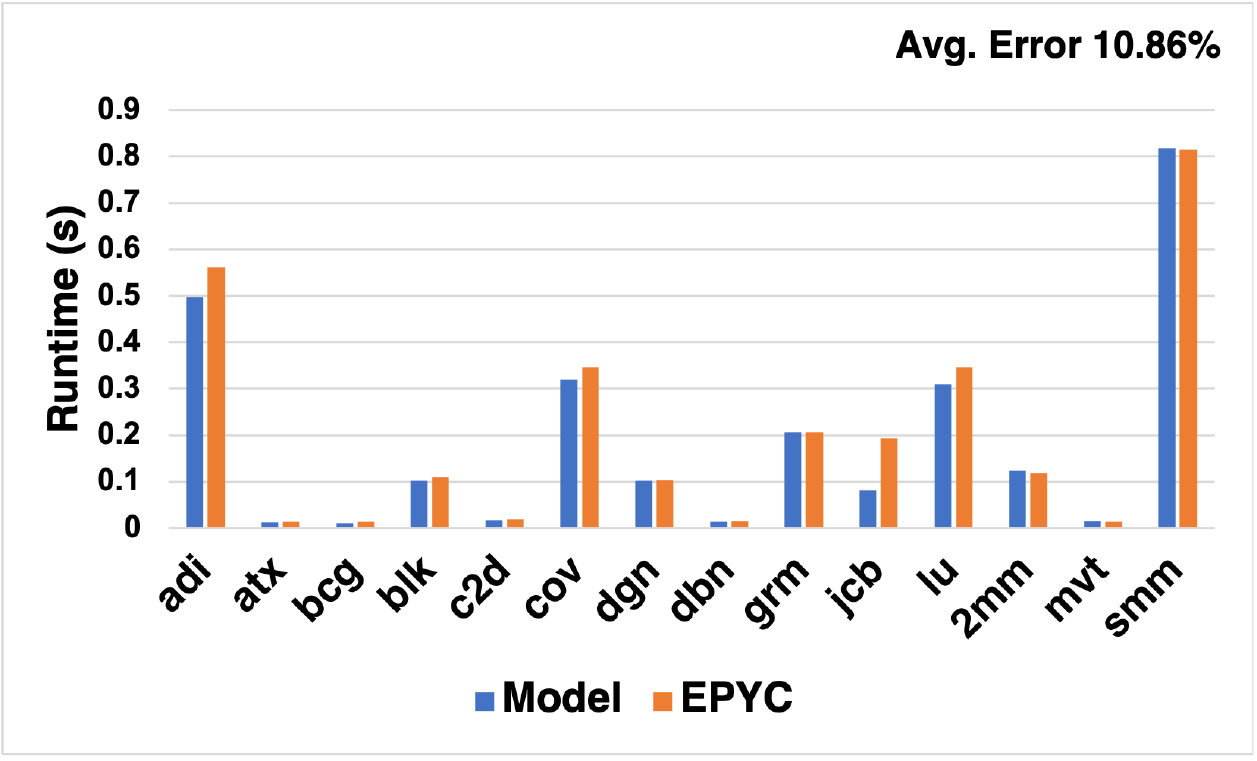}}%
\qquad
\caption{OpenMP applications' parallel sections' runtime comparison on AMD EPYC 7702P with different core configurations.}
\label{fig:results-runtime-epyc}%
\end{figure}

\vspace{-10pt}
\subsection{Runtime Verification}
\label{sec:runtime_veri}
\vspace{-5pt}
We model the processors listed in table~\ref{table:cpus_for_veri} inside PPT to predict runtimes of the benchmark applications. The previous implementation of PPT had parameterized models for single-core CPUs~\cite{chennupati:pads} and GPUs~\cite{arafa-ppt-gpu}. We added support for multicore modeling in PPT. \emph{PPT-Multicore} takes the number of cores, clock frequency, cache sizes, line sizes, associativity, cache latencies, reciprocal throughput of different types of instructions(\emph{e.g.}, integer arithmetic, floating-point arithmetic), data bus width, RAM bandwidth, and RAM latency as hardware parameter inputs. We take the hardware parameter values from the processor manufacturer's website, Agner Fog's instruction latency manual~\cite{agner-instruction}, and 7-CPU website~\cite{7CPU}.

Our model's software parameters include the number of different types of integer arithmetic operations, floating-point arithmetic operations, block size, and total memory operations in bytes for the parallel section of the benchmarks. We use Byfl~\cite{Byfl} to gather these data from the sequential run of benchmarks. To measure $T_{pred}$, we pass these data along with private and shared reuse profiles to PPT using a tasklist. An example tasklist looks like figure~\ref{fig:tasklist}. In the tasklist, we divide the number of different ALU operations by the number of cores/threads to measure $T_{CPU}$ per core. We use the system clock to get runtimes of parallel sections on the real machine.

Figure~\ref{fig:results-runtime-i7} shows the runtime comparison between our model and the Intel Core I7-5960X processor. Runtimes are shown in seconds. The average error rates are 10.78\%, 8.06\%, 8.82\%, and 13.63\% for 1, 2, 4, and 8 core configurations respectively. On Xeon E5-2699 the error rates are 9.93\%, 7.08\%, 7.53\%, 12.41\%, and 16.72\% as shown in figure~\ref{fig:results-runtime-xeon}. We compare the predicted runtimes for AMD EPYC 7702P in figure~\ref{fig:results-runtime-epyc}. The average error rates are 6.81\%, 6.32\%, 6.82\%, 7.34\%, and 10.86\% for 1, 2, 4, 8, and 16 core configurations respectively. Overall we have an average error rate of \textbf{9.08\%} for runtime prediction. As we do not model some modern CPU features like pipelining, speculative execution, branch prediction in the current PPT version, we notice higher error rates in runtime prediction. We also notice that some applications' runtime does not scale with core count. In those cases, our model provides a high error rate. As an example, \emph{Jacobi} has high error for 8 core configuration (36.43\%) as its runtime does not scale much from 4 (runtime 0.128s) to 8 (runtime 0.092s) core count on a Core i7 processor. On Xeon and EPYC, we also notice a similar result with \emph{Jacobi}. Its runtimes are 0.111 seconds and 0.119 seconds for 8 and 16 core configurations on Xeon. On EPYC the runtime of \emph{Jacobi} increases from 0.175s(8 core) to 0.194s(16 core) with increase of core count. As a result, we notice  55.60\% and 58.62\% error rates in Jacobi's runtime prediction on Xeon and EPYC, respectively, for 16 core configurations. This behavior is expected as we divide the number of different CPU operations by core count when we pass the tasklist to PPT as shown in figure~\ref{fig:tasklist}, which is later used for computing $T_{CPU}$ per core.

\vspace{-10pt}
\section{Related Works}
\label{sec:related_work}
\vspace{-5pt}
Recent works in performance modeling can be categorized into two approaches, analytical and simulation based~\cite{survey_evoulution_methods}. Analytical models like~\cite{LogGP,LogP,amdahl-multicore,pypasst-obaida,ExaSAT,hybrid-MPI-OpenMP} deduce mathematical equations to predict the performance. Analytical regression  models~\cite{regres-multi,linear-reg-model} has also been proposed by researchers for multicore performance modeling. Several groups also explored and benchmarked machine learning performance modeling~\cite{ml-based-Dubach,ml-based-Ipek-Predictive,ml-benchmark}. Although these models are fast, they are often less accurate than simulators. Although simulators are slower than analytical model, simulators such as Simics~\cite{func:simics}, SimpleScalar~\cite{func:simplescalar}, SimCore~\cite{func:simcore}, HASE~\cite{func:hase}, Barra~\cite{func:barra}, MSim~\cite{timing:msim}, RSim~\cite{timing:rsim}, SimFlex~\cite{timing:simflex}, and Sniper~\cite{snipersim} have been proposed by the researchers for performance prediction and estimation at various levels of detail.

Simulators such as gem5~\cite{gem5}, MARSS~\cite{marss}, and the Structural Simulation Toolkit~\cite{sst} aim to produce cycle-accurate predictions, which is time-consuming but important for concentrated design space explorations.

Alternatively, recent attempts in program analysis include COMPASS~\cite{compass}, Durango~\cite{durango}, CODES~\cite{CODES} which rely on Aspen~\cite{aspen}. Aspen performance modeling takes source code as input and generates the programs' control flow graph while combining it with analytic modeling aspects.

On the other hand, reuse distance~\cite{Mattson:RD:IBM} analysis is a commonly used method for cache performance prediction~\cite{performance:Beyls:RD:2001,performance:CaBetacaval:2003:ECM,multicore:Fast_and_Accurate_Exploration:Maeda,performance:Sen:2013:ROM,badawy-cal,badawy-ipccc}, cache policy management~\cite{Das:CacheReplacement,C.Management:Duong:2012:ICM,C.Management:Keramidas:2007}, and program locality prediction~\cite{Berg:SS,locality:Ding:2003:PWL,Jiang:RD-Applicable-on-chip,locality:Zhong:2009:PLA}. Researchers have also used and further parallelized graph algorithms to efficiently implement reuse distance analysis~\cite{PARDA:Niu} and proposed analytical modeling and sampling techniques~\cite{Shen:2007:LAU,ppt-amm,chennupati:pads,chennupati:pmbs}. Recently, several research works on multicore and GPU reuse profile analysis has been also published~\cite{Jiang:RD-Applicable-on-chip,Multicore_Reuse_Analytical:Jasmine,Schuff:2010:AMR:1854273.1854286,multicore:stat_multiprocessor_cache:Berg,Multicore-Aware-Derek,Wu-multicore-journal,arafa_ics,arafa_ipccc}

Jiang~\emph{et al.}~\cite{Jiang:RD-Applicable-on-chip} introduced CRD profiles for multicores and provided a probabilistic model to estimate the CRD of each thread. They do not consider invalidation for data locality analysis of private caches.

Wu and Yeung~\cite{Wu-multicore-journal} explored PRD and CRD profiles for performance prediction of loop-based parallel programs. They provided a detailed analysis of the effect of core count on PDR and CRD profiles. They also developed a model for predicting PRD and CRD profiles with core count scaling. The predict the CRD profile with about 90\% accuracy. In other work~\cite{wu-ml-rd-scaling}, they studied the impact of core count and problem size scaling on the program locality's predictability.

Sabarimuthu~\emph{et al.}~\cite{Multicore_Reuse_Analytical:Jasmine} proposed a probabilistic method to calculate the CRD profile of threads sharing a cache and derived coherent reuse profile of each thread considering the effect of cache coherence. They derived the concurrent reuse distance (CRD) profile of each thread, sharing the cache with other threads from the respective thread's private reuse profile.

In prior work to this paper, we~\emph{et al.}~\cite{ppt-sasmm} proposed a probabilistic method to predict PRD and CRD profiles from single-threaded execution trace and introduced \emph{PPT-SASMM}. Although \emph{PPT-SASMM} can measure reuse profiles accurately, the implementation was slow. This paper takes the ideas of \emph{PPT-SASMM} and re-implements it for performance improvement. Compared to PPT-SASMM, our private and shared cache trace generation is much faster. We also employ a tree-based algorithm to measure reuse profiles which makes our approach much faster~\cite{PARDA:Niu}.

Schuff~\emph{et al.}~\cite{Multicore-Aware-Derek} explored reuse distance analysis for shared cache accounting inter-core cache sharing. They also studied PRD profiles considering invalidation-based cache-coherence. They further extended their work to accelerate CRD profile measurement by introducing sampling and parallelization~\cite{Schuff:2010:AMR:1854273.1854286}.

Ding~\emph{et al.}~\cite{Multicore:ding2009a} explored theories and techniques to measure program interaction on multicores and introduced a new footprint theory. They proposed a trace-based model that computes a set of per-thread metrics. They compute these metrics using a single pass over a concurrent execution of a parallel program. Using these metrics, they propose a scalable per-thread data-sharing model. They also propose an irregular thread interleaving model integrated with the data-sharing model.

Kaxiras~\emph{et al.}~\cite{multicore:Kaxiras} proposed statistical techniques from epidemiological screening and polygraph testing for coherence communication prediction in shared-memory multiprocessors.

Almost all of these approaches collect traces at different cache levels from parallel execution of the application. Our approach is different since we collect a trace only once from a sequential execution of the application. This makes our approach more scalable with core count.
\vspace{-10pt}
\section{Conclusion and Future Direction}
\label{sec:conclusion}
\vspace{-5pt}
We introduced \emph{PPT-Multicore} which improves our previous approach for cache hit rate prediction of OpenMP program on multicores and adds runtime prediction of the parallel sections. Given an OpenMP parallelizable program and the modeled multicore parameters, \emph{PPT-Multicore} can accurately predict the runtime of parallel sections. This prediction involves several intermediate results, including private and concurrent reuse profile prediction, latency and throughput of memory accesses, and finally, the runtimes. We validated our model using different benchmark applications from different domains with standard input sets on Intel Core i7-5960X, Intel Zeon E5-2699 v4, and AMD EPYC 7702P processors. The results show that our model can predict the hit rates at different cache levels accurately. Our runtime prediction is also promising, with an overall average error rate of 10.53\%. We plan to model pipelining, speculative execution, prefetching, and other modern computer architecture features of multicores in the future.

\vspace{-5pt}
\begin{acknowledgements}
The authors would like to thank Dr. David Newsom for donating several machines to the PEARL laboratory at NMSU. Some of the experiments in this paper were run on the donated machines. This work is partially supported by Triad National Security, LLC subcontract \#581326. This paper has been approved for unlimited public distribution under LA-UR-21-22749. Any opinions, findings, and/or conclusions expressed in this paper do not necessarily represent the views of the DOE or the U.S. Government.
\end{acknowledgements}
\vspace{-15pt}

%
%

\scriptsize
\bibliographystyle{spmpsci}      


\end{document}